\documentclass[aps, prd, twocolumn, groupedaddress, showpacs, showkeys]{revtex4}
\usepackage{graphicx}
\usepackage{dcolumn}
\usepackage{mathrsfs}
\usepackage{amsmath}
\usepackage{latexsym}

\def\om{\omega}

\def\bx{{\mathbf{x}}}

\newcommand{\ben}{\begin{equation}}
\newcommand{\een}{\end{equation}}
\newcommand{\bea}{\begin{eqnarray}}
\newcommand{\eea}{\end{eqnarray}}
\newcommand{\ba}{\begin{array}}
\newcommand{\ea}{\end{array}}
\newcommand{\bit}{\begin{itemize}}
\newcommand{\eit}{\end{itemize}}

\def\math{\mathsurround 0pt}
\def\oversim#1#2{\lower.5pt\vbox{\baselineskip0pt \lineskip-.5pt
        \ialign{$\math#1\hfil##\hfil$\crcr#2\crcr{\scriptstyle\sim}\crcr}}}

\begin{document}

\title{Numerical Investigations of Oscillons in 2 Dimensions}
\author{Mark Hindmarsh$^{1}$}
\email{m.b.hindmarsh@sussex.ac.uk}
\author{Petja Salmi$^{1,2}$}
\email{salmi@lorentz.leidenuniv.nl}
\affiliation{$^{1}$Department of Physics \& Astronomy, 
University of Sussex, Brighton BN1 9QH, UK\\
                   $^{2}$Lorentz Institute of Theoretical Physics, 
University of Leiden, The Netherlands}
\date{\today}

\begin{abstract}
Oscillons, extremely long-living localized oscillations 
of a scalar field, are studied in theories with quartic and 
sine-Gordon potentials in two spatial dimensions. 
We present qualitative results concentrating largely on a study in
frequency space via Fourier analysis of oscillations. 
Oscillations take place at a fundamental frequency just below 
the threshold for the production of radiation, with exponentially 
suppressed harmonics. The time evolution of the 
oscillation frequency points indirectly to a life time of at 
least $10^7$ oscillations.
We study also elliptical perturbations of the oscillon, which are shown to 
decay.
We finish by presenting results for boosted and collided oscillons, 
which  point to a surprising persistence and soliton-like behaviour.
\end{abstract}


\maketitle

\section{Introduction}

There have been many studies on static non-dissipative solutions 
of classical field theories when the stability of the solutions is guaranteed 
due to a conservation of a topological charge 
(see~\cite{'tHooft:1974qc,Polyakov:1974ek}) 
or conserved particle number~\cite{Coleman:1985ki} and there exist examples 
where exact solutions are known and can be written explicitly 
in a closed form (see e.g.~\cite{Prasad:1975kr,Theodorakis:2000bz}).
However, energy can be stored in localised oscillations of a field for a 
very long time and in this situation there is no obvious 
conservation law to explain the (meta-)stability of the configuration.
In 1+1 dimensions 'breather' solutions are known 
and can be written down in a closed form \cite{Rajaraman}. 
In higher dimensions, 
the long life time of certain oscillations was pointed out 30 years ago 
in~\cite{Bogolyubsky:1976nx,Bogolyubsky:1976sc} 
and then re-discovered~\cite{Gleiser:1993pt} when the 
dynamics of first order phase transitions and bubble nucleation was studied. 
More extended investigations were carried out in~\cite{Copeland:1995fq}, 
where conditions, like the size of the initial bubble, 
for long living oscillations were examined.
In~\cite{Bogolyubsky:1976sc} oscillating objects were 
called pulsons, we adopt hereafter the definition {\it {oscillon}} according 
to~\cite{Copeland:1995fq}.

There is still not a satisfactory explanation for the long lifetime of 
oscillons. 
In~\cite{Kasuya:2002zs} it was suggested that the longevity of 
oscillations could be understand within the framework of adiabatic 
invariance, i.e. there were a conserved adiabatic charge explaining 
the metastability of oscillons, called I-balls by the 
authors~\cite{Kasuya:2002zs}. 
However, this approach assumes negligible gradient energy and 
non-linearity, i.e. quadratic potential. 
While the adiabatic invariance, in spite of its restrictions, 
provides an analytic approach to explain the long, 
but finite life time, some authors have suggested, 
based on numerical studies, that the life time 
could be arbitrarily long~\cite{Watkins,Honda:2001xg}, and even 
a Lyapunov exponent was suggested to govern the 
power law of oscillons life time~\cite{Honda:2001xg}.

The possible effects of oscillons on the dynamics of 
first order phase transitions has been studied 
e.g. in~\cite{Gleiser:2003uu} 
(for creation of long-living quasilumps from two bubble 
collisions see~\cite{Dymnikova:2000dy}).
It was pointed out by Riotto~\cite{Riotto:1995yy} that 
at the electroweak scale thermal fluctuations are too weak to 
create oscillons and therefore they cannot play a role in 
first order electroweak phase transition. However, 
a very recent study~\cite{Gleiser:2006te} exploited the 
possibility that oscillons would speed up the phase transition 
in the context of old inflation. 
It is natural, that oscillons are not persistent 
in a hot heat bath, but they can be long living in a 
weak enough thermal environment~\cite{Gleiser:1996jb}.
Principally, the necessary conditions of potentials 
to permit oscillons can be fulfilled also in 
higher dimensional models~\cite{Gleiser:2004an}, but on the other hand 
it should be noticed that stable or extremely long living oscillons 
(called pseudobreathers by the authors) were obtained in sine-Gordon 
model only in two spatial dimensions~\cite{Piette:1997hf}.

In addition to~\cite{Gleiser:2006te} there has been recent 
interest in oscillons and they have been found 
in some realistic models of great importance. 
In~\cite{Farhi:2005rz} authors reported that with a particular, 
experimentally relevant, ratio between Higgs and W boson masses, 
there exists an oscillon in an SU(2) model that is essentially
the bosonic electroweak sector of the Standard Model, 
at zero weak mixing angle. 
An oscillon in the Standard Model could have significant 
consequences e.g. for baryogenesis.
Another study~\cite{Broadhead:2005hn} found oscillon 
formation after supersymmetric hybrid inflation. 
The authors worked within the very successful D-term 
inflationary scenario and concluded that for large 
enough, but fully realistic, couplings in the model, 
oscillons (called non topological solitons in~\cite{Broadhead:2005hn}), 
form in large numbers and even dominate the energy density in the 
post-inflationary era (also in~\cite{Copeland:2002ku} energy concentrations, 
often along the domain walls, were reported in 
simulations of tachyonic preheating).

Our study here was motivated by creation of oscillons from the collapsing 
domains in $\phi^4$ theory and in the sine-Gordon model, 
a topic we hope to revisit in another study.
Here our investigations concentrate on the properties of oscillons 
when they are 
created with a Gaussian initial ansatz instead of as a consequence 
of the evolution after 
random initial conditions. However, we try to draw the connection by 
studying the boosted oscillons in the end of this paper. 
Before that we present results from a stationary set-up using 
Fourier analysis to determine the time evolution 
of the oscillation frequency when oscillon is created using a 
spherically symmetric initial ansatz and studies 
of the collapse to spherical symmetry for an elliptic ansatz.

\section{The models and numerical set-up}
\label{s:Models}

The Lagrangian of a model for a single real scalar field~$\phi$ 
in the presence of a potential $V$ is given by
\begin{eqnarray}
  \mathscr{L} =
 \frac{1}{2} {\partial}_{\mu} \phi {\partial}^{\mu} \phi - V(\phi), 
 \label{lagrangian}
 \end{eqnarray}
and the equation of motion thus reads
\begin{eqnarray}
\ddot{\phi}-\nabla^{2}\phi + V'(\phi)=0. 
 \label{eqm}
\end{eqnarray}
We study here two potentials with a discrete symmetry: 
degenerate double-well quartic potential
\begin{eqnarray}
V(\phi)=\frac{1}{4} \lambda (\phi^2-v^2)^2
 \label{quartic} 
\end{eqnarray}
and sine-Gordon potential
\begin{eqnarray}
V(\phi)= \frac{\Lambda^4}{\alpha^2} \big( 1 - \cos (\alpha \phi) \big).
  \label{sine-Gordon}
\end{eqnarray}
Scaling out the vacuum expectation value and couplings these can be written
\begin{eqnarray}
V(\phi)=\frac{1}{4} (\phi^2 - 1 )^2 
 \label{scaledquartic}
\end{eqnarray}
and 
\begin{eqnarray}
V(\phi)=  \frac{1}{\pi^2}\big( 1 + \cos (\pi \phi) \big) 
  \label{scaled-sine-Gordon}
\end{eqnarray}
so that both~(\ref{scaledquartic}) and~(\ref{scaled-sine-Gordon}) have 
minima at $\phi = \pm 1$ and a local maximum at $\phi = 0$.

The field equation is evolved on a two-dimensional lattice 
with periodic boundary conditions using a leapfrog update and 
a three-point spatial Laplacian accurate to O($dx^2$).
The lattice spacing for the data shown is set to be $dx=0.25$ and 
the time step $dt=0.05$. With that choice the 
fluctuations in total energy on the lattice are less than 0.2\% 
over $9 \times 10^{7}$ iteration steps. 
We tested our method by reducing both $dx$ and $dt$ from the above mentioned 
choice without observing significant difference in the quantities of main 
interest here.
Unless otherwise stated the simulations for the data shown were carried out
in $800^2$ lattices. We choose periodic boundary conditions 
rather than absorbing ones 
because we wish to explore the stability of the oscillons to the 
small perturbations 
from any radiation emitted, and because we wish to allow them 
to move without hitting any boundaries.

\section{Properties}

We start with a Gaussian ansatz 
\begin{eqnarray}
\phi(r) = v( 1 - A \exp(-r^2/r_{0}^2))
 \label{gaussian-ansatz}
\end{eqnarray}
where $r$ is the distance to the center of an oscillon 
$r=(x_{1}^2+x_{2}^2)^{1/2}$ and the width of the distribution was set to be 
$r_0 \simeq 2.9$ (in units of  
$(\sqrt{\lambda}\, v )^{-1}$), suggested an optimal choice for a long living 
oscillon in~\cite{Copeland:1995fq}. 
Earlier studies have established the sensitivity both to the spatial 
size $r_{0}$ 
and the amplitude of the deviation from vacuum.
For the data shown amplitude $A=1$ so that the center of 
the oscillon is located in the local maximum of the potential. 
This choice sets the initial deviation from vacuum to be drastic 
and ensures that we observe the non-linear features of 
the theory and not a small, linear perturbation around the vacuum.

It should be immediately noted that the Gaussian ansatz does not provide the 
right description of radiation far from oscillon core. This 
can be seen by studying a small, radially symmetric spatial perturbation 
$\varphi$ around the 
vacuum. For quartic potential~(\ref{quartic}) substitution $\phi=v-\varphi(r)$ 
into~(\ref{eqm}) leads in lowest order to
\begin{eqnarray}
\frac{{\rm d}^2 {\varphi(r)}}{{\rm d}^2 r} + 
\frac{1}{r}\frac{{\rm d}{\varphi (r)}}{{\rm d}r}+ m^2 \varphi(r)=0,
 \label{bessel}
\end{eqnarray}
where $m^2=2\lambda v^2$. This Bessel equation has solution 
$\varphi=C_1 \cdot J_{0}(mr) +C_2 \cdot Y_{0}(mr)$, where $J_{\nu}$ and 
$Y_{\nu}$ are the 
Bessel functions of the first and second kind, respectively.  
Both are oscillatory with an amplitude decaying 
asymptotically as $r^{-1/2}$, thus much more slowly 
than~(\ref{gaussian-ansatz}). Repeating the calculation for 
sine-Gordon potential~(\ref{sine-Gordon}) yields 
the same result with $m=\Lambda^2$.
The parameter $m$ defines a mass in the theory as $m^2 = V''(\phi)$, 
where $\phi$ is at the minimum of the potential.

Watkins suggested in~\cite{Watkins} the following 
spherically symmetric periodic trial solution (see also~\cite{Honda:2001xg}) 
\begin{eqnarray}
\phi(r,t) = \sum_{n=0}^{\infty} f_{n}(r) \cos(n \omega t) \,,
 \label{solution-ansatz}
\end{eqnarray}
with $\omega$ a free parameter.
Studying the coupled ODE's for $f_{n}$ it was noticed that 
$f_{n} \rightarrow 0$ fast as $n$ increases, so that for $n \ge 5$ they 
are negligible. 
It was also found that solutions existed only for $\om < m$, going some way to 
explaining the stability: the fundamental mode of oscillation is 
not quite fast enough 
to excite outgoing modes.
We bear this in mind when discussing our numerical 
results for $\phi^4$ and sine-Gordon potentials.

\subsection{$\phi^4$ potential}

\begin{figure}
\begin{center}

\includegraphics[width=0.94\hsize]{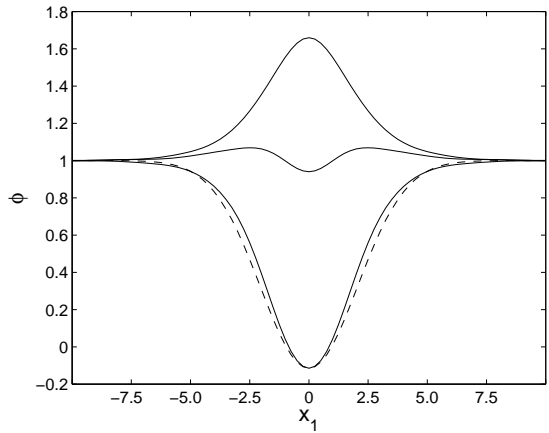}
  \caption{\label{f:oscillon_profile} Oscillon profile at the extrema 
(above and below the vacuum expectation value) and at 
crossing the minimum of the potential in the 2D $\phi^4$ theory. 
Dashed line shows a Gaussian form with the same amplitude 
and the width $r_0$ of the initial profile.} 

\end{center}
\end{figure}

\begin{figure}
\begin{center}

\includegraphics[width=0.94\hsize]{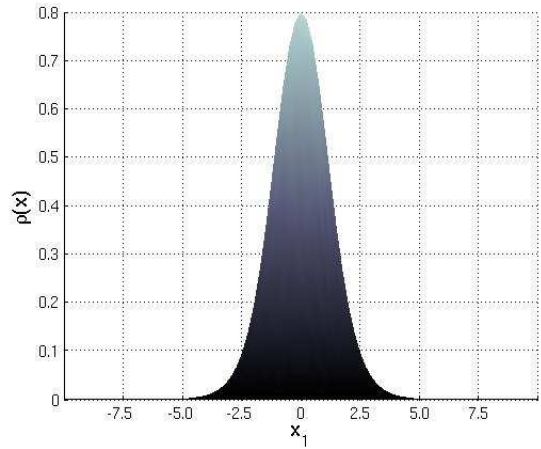}
\includegraphics[width=0.94\hsize]{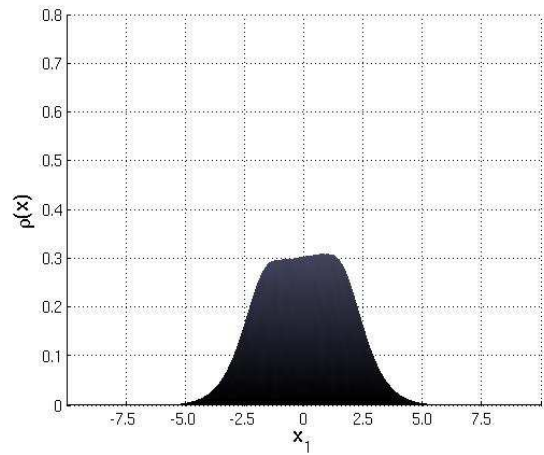}

  \caption{\label{f:energy-density} Energy density $\rho(\boldsymbol{x})$ 
of an oscillon in $\phi^4$ theory at the moment of crossing 
the minimum of the potential (above) and at the 
moment of maximum excursion (below), corresponding the field $\phi$ shown 
by the lowest solid line 
in Figure~\ref{f:oscillon_profile}.}

\end{center}
\end{figure}

\begin{figure}
\begin{center}

\includegraphics[width=0.94\hsize]{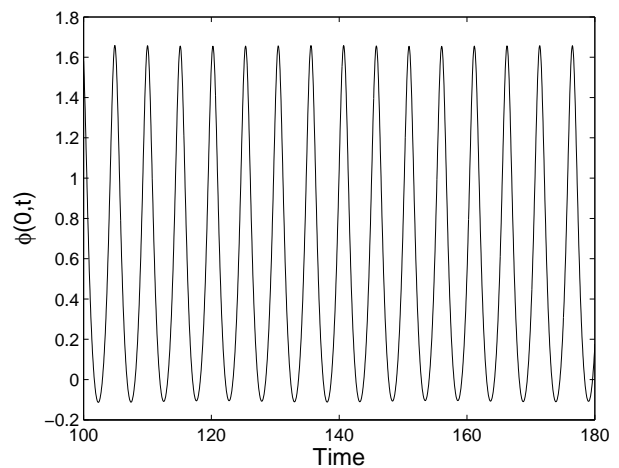}
  \caption{\label{f:phi4CentreField} Field at the centre of 
the $\phi^4$ oscillon $\phi(t,0)$ as a function of time $t$. }

\end{center}
\end{figure}

\begin{figure}
\begin{center}

\includegraphics[width=0.94\hsize]{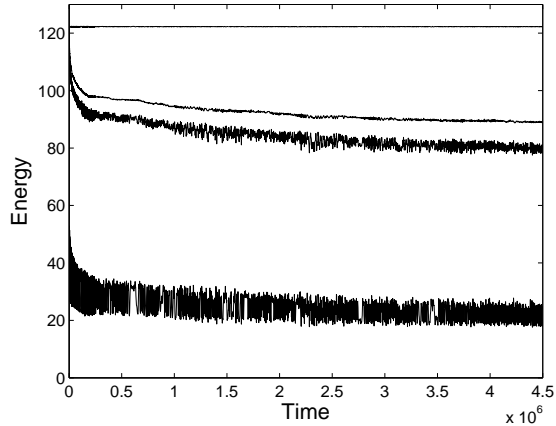}
  \caption{\label{f:phi4energyshells} Total energy in the lattice and energy 
inside the consecutive shells with radius $r=2.5\,r_0$,~$1.5\,r_0$ 
and $0.5\,r_0$ around the center of the $\phi^4$ oscillon (from top to bottom) 
as a function of time~$t$. }

\end{center}
\end{figure}

\begin{figure}
\begin{center}

\includegraphics[width=0.94\hsize]{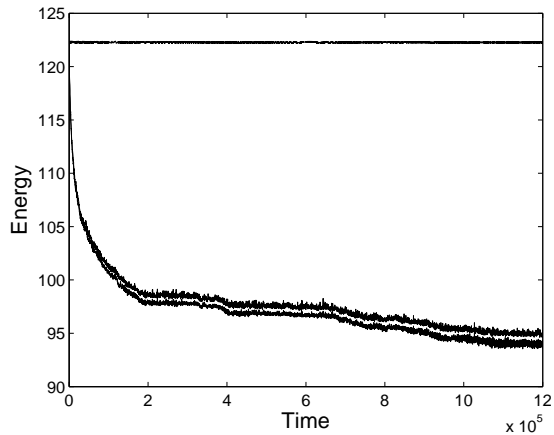}
  \caption{\label{f:phi4energyshells2} Detail from 
  Figure~\ref{f:phi4energyshells}, showing the total energy in the lattice 
and the energy inside shells with radius $r=5\,r_0$ and $2.5\,r_0$ 
around the center of the $\phi^4$ oscillon (from top to bottom).
 }
\end{center}
\end{figure}

\begin{figure}
\begin{center}

\includegraphics[width=0.94\hsize]{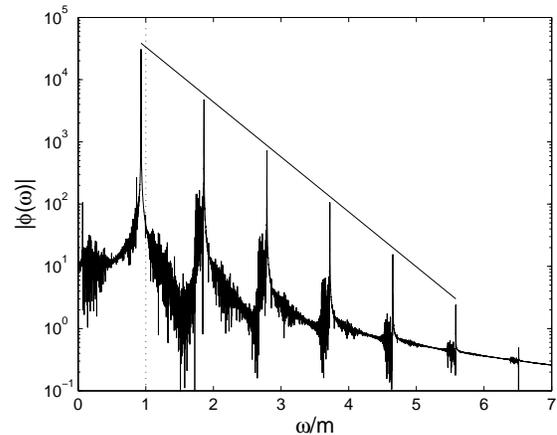}
  \caption{\label{f:phi4Spectrum} Power spectrum of field at 
centre of $\phi^4$ 
  oscillon $|\phi(\om)|$ taken in time 
interval $2.5\cdot 10^5< t < 2.55 \cdot 10^5$.
Solid line is a guide to eye for an exponential fit to the amplitudes of 
the first six peaks, 
$\exp(- a \omega/m)$ with a slope $a=2.03$. 
Vertical dotted line shows the radiation frequency $\omega=m$. 
}

\end{center}
\end{figure}

\begin{figure}
\begin{center}

\includegraphics[width=0.94\hsize]{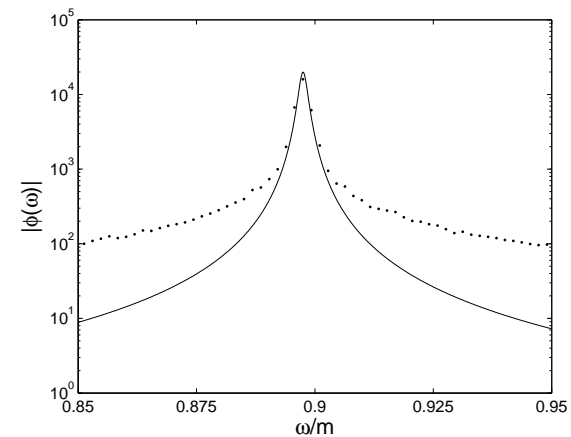}
  \caption{\label{f:phi4BWfit} Fit (solid line) to Breit-Wigner 
form of first peak (dots) $\phi^4$ 
  oscillon taken in time interval $10^4< t < 1.25 \cdot 10^4$: 
$\omega_{0}=0.8975$, $\Gamma = 0.0020$ and $K=0.02$.}

\end{center}
\end{figure}
\begin{figure}
\begin{center}
\includegraphics[width=0.94\hsize]{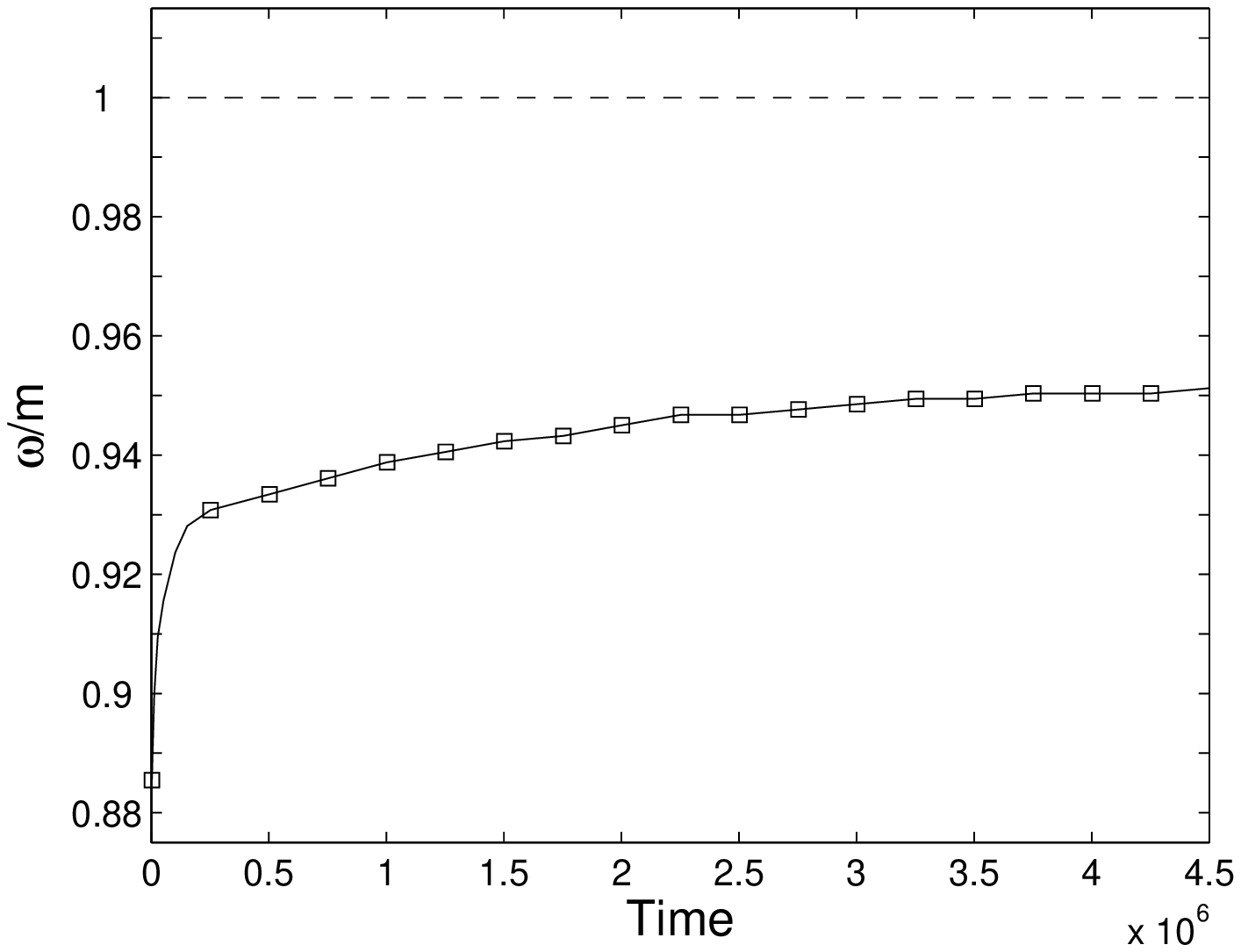}
  \caption{\label{f:phi4peakfreq} Oscillation frequency corresponding to 
the location of the highest peak as a function of 
time for $\phi^4$ oscillon. The precision of the measurement of 
the frequency is $0.002$ at the scale of $\omega /m$ due to the limited time 
interval the Fourier transformation is made. Some data points 
close to the frequency axis are not marked with $\Box$ for clarity. 
Dashed line shows the radiation frequency $\omega = m$.} 
\end{center}
\end{figure}

The shape of the oscillon profile in $\phi^4$ theory is shown 
in Figure~\ref{f:oscillon_profile}. Deviation from the Gaussian form is 
clearly visible, but not drastic. Figure~\ref{f:energy-density} 
illustrates the broadening of the energy 
density $\rho(\boldsymbol{x})$ in the extremum and contraction to a highly 
concentrated spike as oscillation crosses the minima and kinetic 
energy dominates the density. 
The value of the field at the centre of the oscillon as a function of time is 
plotted in the Figure~\ref{f:phi4CentreField}. 
An oscillation over one period is not symmetric, as the field makes 
larger excursion and changes more slowly where the potential is flatter 
whereas it changes rapidly in the region of steeper potential. 
Variation of the amplitude does not become apparent 
over this short period of time.

The total energy and the energy inside shells of radius 
$r=0.5r_0,\,1.5r_0$~and~$2.5r_0$ around the centre of the oscillon 
over the complete simulation time ($4.5 \times 10^6$) 
are shown in Figure~\ref{f:phi4energyshells}. Size $r=0.5r_0$ shows 
large fluctuations depending on the phase of the oscillation, 
which can be well understood on the basis of the behaviour 
of $\rho(\boldsymbol{x})$ over a period (Figure~\ref{f:energy-density}).
On the contrary shell $r=2.5r_0$ shows relatively thin line and provides 
a good estimator for the energy of an oscillon.
This is further confirmed by comparing the energy inside shells of 
radius $r=2.5r_0$~and~$5r_0$. They track each other well as can be 
seen in Figure~\ref{f:phi4energyshells2}, where they are plotted 
for a shorter time interval from the beginning of the simulation. 
Figure~\ref{f:phi4energyshells2} shows also clearly a period in the 
beginning when oscillon radiates energy rapidly. After that the 
rate of energy loss is much weaker and the data suggests some 
evidence for the existence of flat plateaus 
where energy stays at constant value and abrupt drops between them.
These plateaus could be interpreted as series of metastable states 
(see also~\cite{Watkins}). 
In the end of the simulation more than $70\%$ of the energy is still 
localised in area covering less than $0.5\%$ of the lattice.

The power spectrum of oscillations was studied 
by performing Fourier transforms of $\phi(t,0)$ in consecutive time intervals. 
The length of the interval for the data shown is 5000 in natural units, 
which amounts just over $0.1$\% of the total length 
covered by the simulation. An 
interval of that length includes approximately $10^3$ oscillations. 
A typical example of the power spectrum 
is shown in Figure~\ref{f:phi4Spectrum}. There are very distinctive 
peaks that rise several 
orders of magnitude higher than the backgound between them. 
The first, highest peak just below frequency $\omega = m$ indicates 
the oscillation frequency, the other peaks are located at integer 
multiplies of that. Up to seven peaks can be identified in the power spectra.

There is a significant change in the shape of the peaks during the time 
evolution. In the beginning, when oscillon radiates and the core loses 
energy reasonably fast, peaks are fairly broad, but later on when 
oscillon 'stabilises' they become extremely narrow. During the early 
evolution it is possible to make a fit to Breit-Wigner formula
\begin{eqnarray}
\sigma(\omega) = \frac{K}{(\omega-\omega_{0})^2 + (\Gamma/2)^2} \, ,
 \label{breit-wigner}
\end{eqnarray}
where $\omega_{0}$ is now the peak frequency. A fit to the first peak is 
shown in Figure~\ref{f:phi4BWfit}. It cannot be 
considered very accurate, but it gives an order of magnitude estimate of 
the decay width $\Gamma$ during the era oscillon radiates its energy strongly. 
Later on the peaks have no width in the restricted time interval 
where the Fourier transformation is made (our choice for the 
length of the interval is fairly optimal as longer intervals 
tend to reveal a shift of the peak frequency, not improve the width).

The amplitudes of the peaks in Figure~\ref{f:phi4Spectrum} obey clearly 
exponential decay law as a function of the frequency $\omega$. 
The deviations from that at the level shown in the figure are most likely 
once again due to the limited accuracy of discrete Fourier transformation. 
The data also points out a low frequency peak 
(just next to $|\phi(\omega,0)|$ axis), with frequency approximately $0.06$. 
There is a slight variation of the amplitude, a beat, with the 
corresponding period around $100$ in time units 
(though not present in Figure~\ref{f:phi4CentreField}), which we believe is 
the cause of the structure.

Not only the shape of the peaks but also their location 
changes during the time evolution.
There exists a strong correlation between the energy in the oscillon 
(shell radius $r=2.5r_0$ in Figure~\ref{f:phi4energyshells}) 
and the oscillation frequency, i.e. location of the highest peak, 
which is shown in Figure~\ref{f:phi4peakfreq} as a function of time. 
During the early period when the oscillon radiates strongly, 
the frequency increases rapidly, but then 
the growth slows down drastically as the rate of energy loss becomes tiny.
The time evolution of the oscillation frequency is the key to predict 
the life time of an oscillon: as long as the frequency 
stays below radiation frequency, oscillon cannot directly radiate 
all of its energy and disappear. As the rate of energy loss decreases, 
the increase in frequency slows down. Turning it other way round this 
is in agreement with Watkins~\cite{Watkins} 
who reported the radiation rate decrease as $\omega /m$ increases, 
with a minimum at $\omega /m \simeq 0.97$. 

Though we have not evolved oscillons in our simulation longer than 
$4.5 \times 10^6$ time units, we can give bounds on the life time 
of oscillon on the basis 
of the evolution of the oscillation frequency. Even a linear fit to 
second half of the points in Figure~\ref{f:phi4peakfreq} suggests 
that the radiation frequency is not reached before a time 
of a few times $10^7$, or about $10^7$ oscillations. 
Similar lifetimes were reported 
in~\cite{Gleiser:1999tj} where lattices with a technique of adiabatic 
damping were exploited. However, the lifetime can be much 
longer as the slope of the line in Figure~\ref{f:phi4peakfreq} is 
clearly flattening out as well as the strongly decreasing radiation 
rate as a function of frequency reported in~\cite{Watkins}.

Setting the original amplitude $A$ too high (e.g. $A=10$ to the 
direction where the potential is steeper) will not lead 
to an oscillon, the initial concentration of energy does not stay 
localised but spreads rapidly. 
Starting with a configuration that has more energy than the one 
shown in Figure~\ref{f:oscillon_profile} can still evolve to an 
oscillon but with a modulated amplitude. This effect becomes most apparent in 
the case of sine-Gordon potential.

\subsection{Sine-Gordon potential}

\begin{figure}
\begin{center}

\includegraphics[width=0.94\hsize]{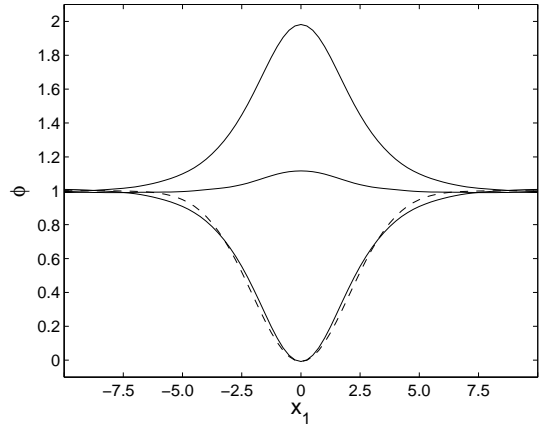}
  \caption{\label{f:axiton_profile} Oscillon profile at the extrema 
(above and below the vacuum expectation value) and at 
crossing the minimum of the potential in the 2D sine-Gordon theory. 
Dashed line shows a Gaussian form with the same amplitude 
and the width $r_0$ of the initial profile.} 

\end{center}
\end{figure}

\begin{figure}
\begin{center}

\includegraphics[width=0.94\hsize]{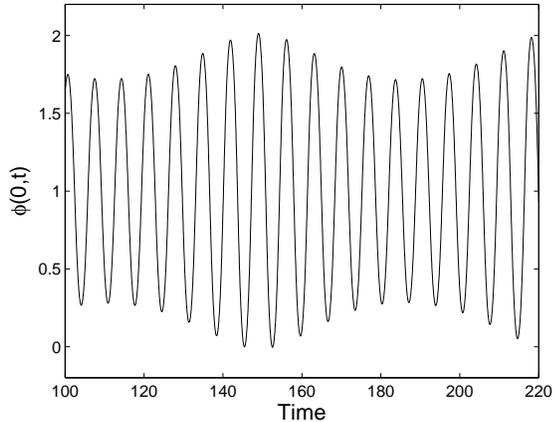}
  \caption{\label{f:SGCentreField} Field at the centre of the 
sine-Gordon oscillon ($\phi(t,0)$) as a function of time $t$.}

\end{center}
\end{figure}

\begin{figure}
\begin{center}

\includegraphics[width=0.94\hsize]{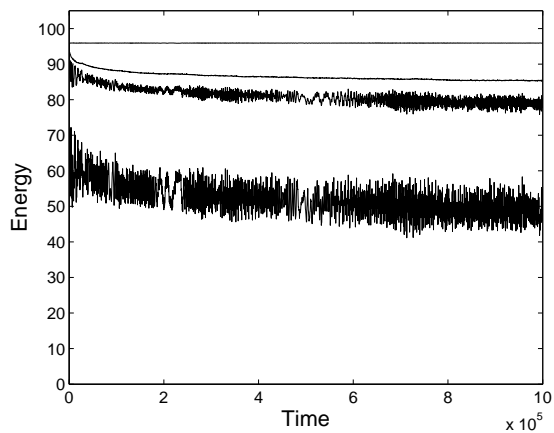}
  \caption{\label{f:SGenergyshells} Total energy in the lattice and energy 
inside the consecutive shells with radius $r=5.0\,r_0$,~$2.0\,r_0$ 
and $1.0\,r_0$ around the center of the sine-Gordon oscillon 
(from top to bottom) as a function of time~$t$. }

\end{center}
\end{figure}

\begin{figure}
\begin{center}

\includegraphics[width=0.94\hsize]{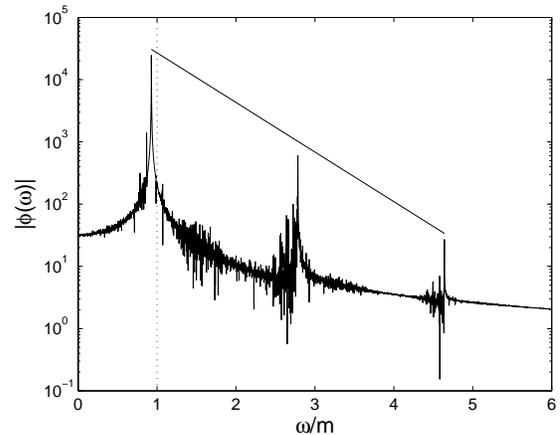}
 
  \caption{\label{f:SGSpectrum} Power spectrum of field at 
centre of sine-Gordon oscillon $|\phi(\om)|$ taken in time 
interval $1.5\cdot 10^5< t < 1.55 \cdot 10^5$.
Solid line is guide to eye for an exponential fit to peak amplitudes, 
$\exp(- a \omega/m)$ with a slope $a=1.84$. 
Vertical dotted line shows the radiation frequency $\omega=m$.
}

\end{center}
\end{figure}

\begin{figure}
\begin{center}

\includegraphics[width=0.94\hsize]{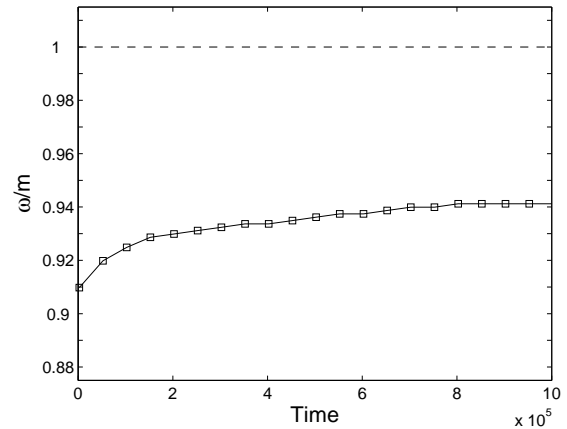}
  \caption{\label{f:SGpeakfreq} Oscillation frequency as a function of 
time for sine-Gordon oscillon. The precision of the measurement of 
the frequency is $0.003$ at the scale of $\omega /m$. 
Dashed line shows the radiation frequency $\omega = m $.}

\end{center}
\end{figure}

Persistent oscillations in radially symmetric sine-Gordon equation 
were studied in~\cite{Piette:1997hf} 
and extreme sensitivity both to the form and the amplitude of 
the initial deviation from the vacuum were observed. A complicated 
initial profile was used to obtain a minimally radiating pseudobreather. 
We have not been able to create a 'minimal' oscillon with a Gaussian 
initial ansatz - though the profile of an oscillon in sine-Gordon potential 
is also fairly close to Gaussian (Figure~\ref{f:axiton_profile}) 
oscillations have modulated amplitude (Figure~\ref{f:SGCentreField}) 
the period of which depends on the initially set amplitude $A$.
Figure~\ref{f:SGenergyshells} shows total energy in the lattice 
and energy inside shells of several radii around the centre of the oscillon 
in a simulation that spans the evolution up to time $10^6$.
There is no apparent evidence for energy plateaus in 
Figure~\ref{f:SGenergyshells}.

The same choice of the interval for Fourier transformation was used 
as in $\phi^4$~theory that now corresponds $0.5$\% of the total time 
of the simulation.
The study in frequency space shows similar basic 
features as in $\phi^4$ theory 
with some significant differences. 
The sine-Gordon potential~(\ref{sine-Gordon}) is symmetric around minima in 
contrast to quartic potential~(\ref{quartic}). 
Due to this symmetry there are no even 
harmonic peaks in the power spectrum of sine-Gordon 
oscillon as can be seen in Figure~\ref{f:SGSpectrum} and 
at very best only first four 
peaks were visible in the power spectrum. As the amplitudes of 
the peaks decrease exponentially also here, 
but only half of them are present compared 
to Figure~\ref{f:phi4Spectrum} of $\phi^4$ theory, this suggests 
that the ansatz~(\ref{solution-ansatz}) would be a 
particularly good approximation for an oscillon in sine-Gordon potential, or 
potentially in any other symmetric potential that allows oscillons. 
There is no sign of low frequency structure in Figure~\ref{f:SGSpectrum} 
in contrast to Figure~\ref{f:phi4Spectrum}.

The time evolution of the frequency of the first peak in power spectrum of 
sine-Gordon oscillon is shown in Figure~\ref{f:SGpeakfreq} from a 
simulation that was evolved $10^6$ time units. 
The oscillation frequency starts 
initially closer to the lowest radiation frequency compared to the quartic 
potential, but there is no drastic increase either, 
which may be due to the weaker coupling to radiative modes as 
even multiples of the oscillation frequency are absent.
Also here the increase in the oscillation frequency slows down 
in the course of time and even linear 
extrapolation yields a life-time estimate, i.e. intersection with the 
radiation frequency, around $10^7$ time units, but it could be much larger, 
especially as the study of the power spectrum suggests weaker radiative modes.

\section{Elliptical oscillons}

\begin{figure}
\begin{center}

\includegraphics[width=0.94\hsize]{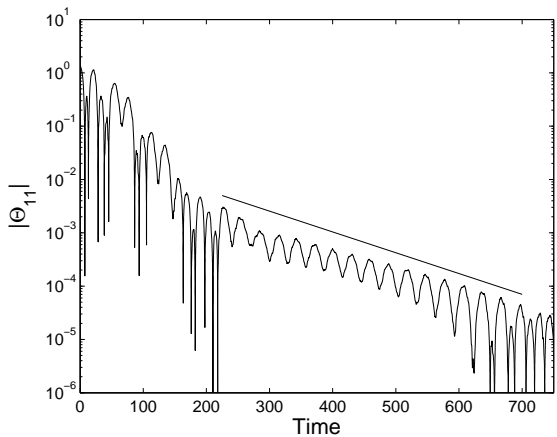}
  \caption{\label{f:phi4ellipticity} Ellipticity, $|\Theta_{11}|$, 
as a function of time in the 2D $\phi^4$ theory. Straight, solid line is 
guide to eye of an exponential fit, $\exp(-bt)$ with a slope $b=0.009$.}

\end{center}
\end{figure}

\begin{figure}
\begin{center}

\includegraphics[width=0.94\hsize]{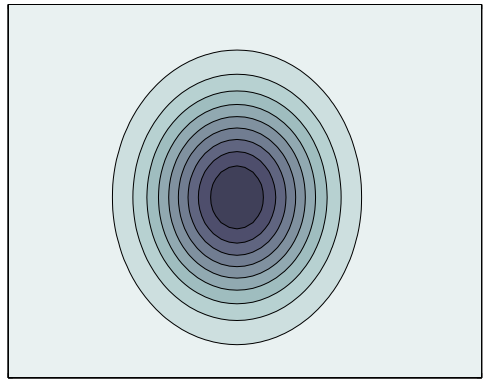}
\includegraphics[width=0.94\hsize]{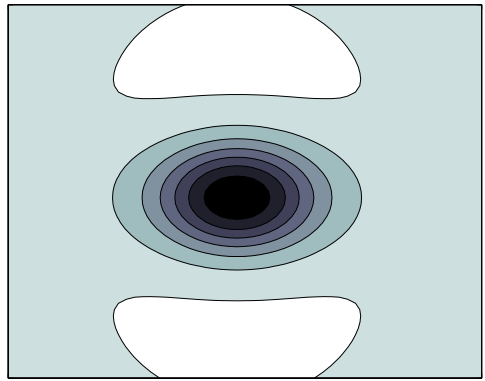}

 \caption{\label{f:phi4elliptical} Contours of $\phi(\bx)$ for 
the initial elliptic 
Gaussian profile and after two oscillations (time~$=10.6$) 
in the 2D $\phi^4$ theory.}

\end{center}
\end{figure}

\begin{figure}
\begin{center}

\includegraphics[width=0.94\hsize]{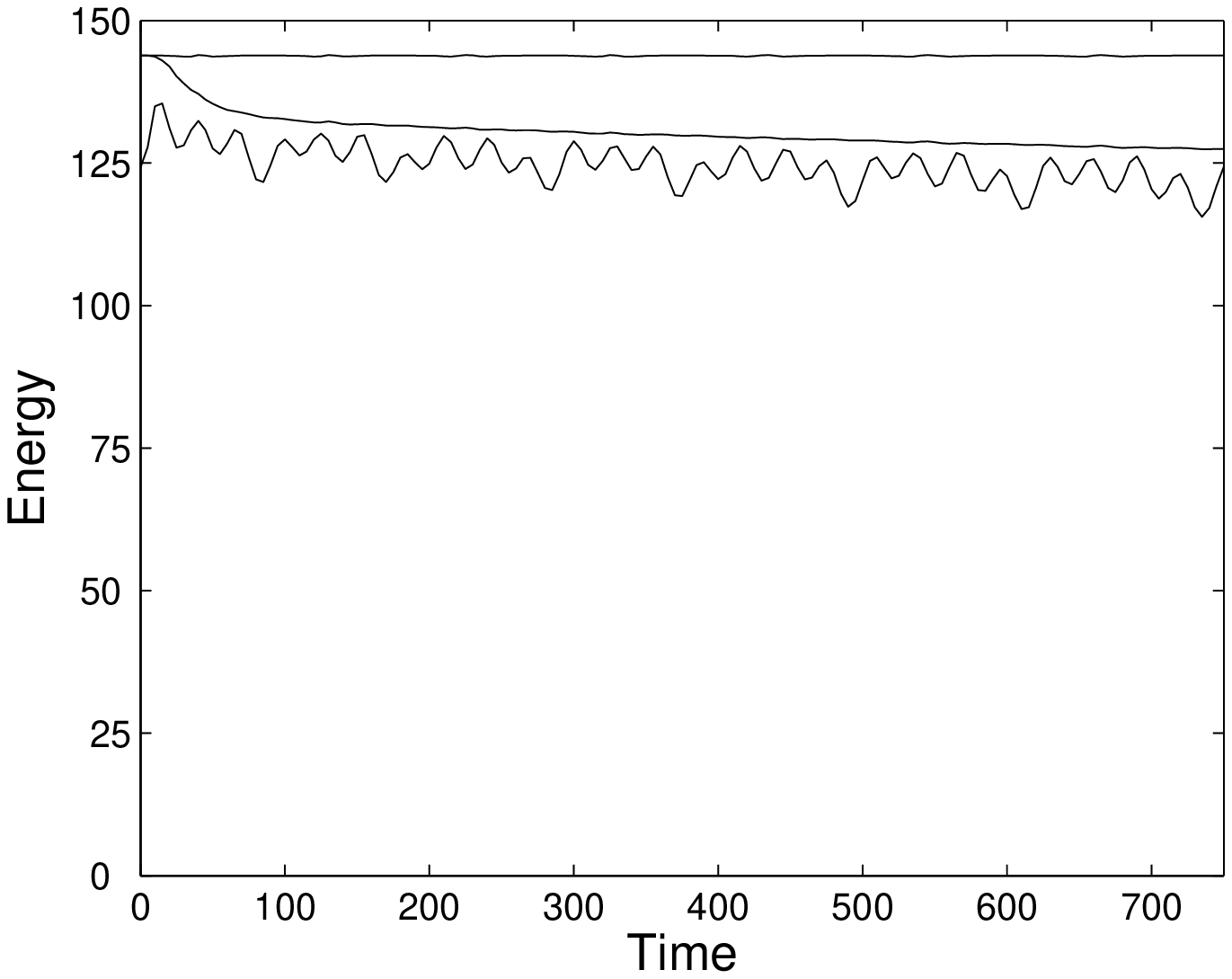}
  \caption{\label{f:ellipticalenergyshells} Total energy and energy 
inside shells of radius $r=2.5\, r_0$ and $r=1.0\,r_0$ around the center of 
$\phi^4$ oscillon as a function of time (from top to bottom). 
Here $r_0=3.9$ is the initial width of the major axis.}

\end{center}
\end{figure}

\begin{figure}
\begin{center}

\includegraphics[width=0.94\hsize]{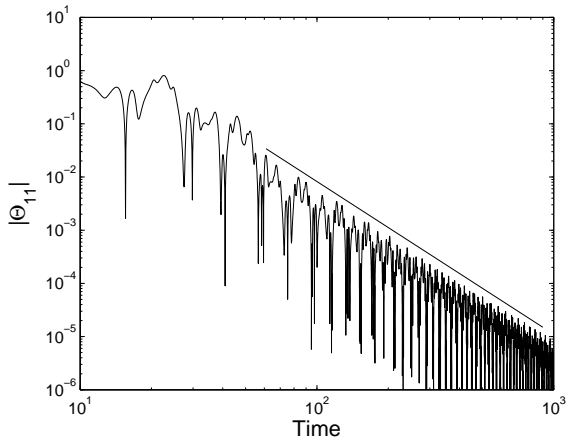}
  \caption{\label{f:SGellipticity} Ellipticity, $|\Theta_{11}|$, 
as a function of time in the 2D sine-Gordon model. Straight, solid line is 
guide to eye of a power law fit, $t^{-\delta}$ 
with a slope $\delta=2.87$.}

\end{center}
\end{figure}

\begin{figure}
\begin{center}

\includegraphics[width=0.94\hsize]{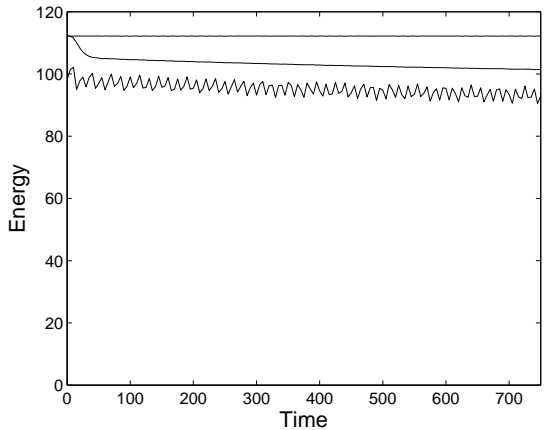}
  \caption{\label{f:ellipticalenergyshellsSG} Total energy and energy 
inside shells of radius $r=2.5\, r_0$ and $r=1.0\,r_0$ around the center of 
sine-Gordon oscillon as a function of time (from top to bottom). 
Here $r_0=3.9$ is the initial width of the major axis.}

\end{center}
\end{figure}

The relative ease to create an oscillon together with the energy shells 
pointing to the existence of successive plateaus and thus a series of 
metastable states gives rise to the question if the field configuration 
of an oscillon is unique. A very modest approach to the uniqueness 
is a study of an initial configuration which is not spherically symmetric.
As our numerical method was not restricted to spherical symmetry 
a Gaussian ansatz with different width 
in $x_1$ and $x_2$ direction was used.
Oscillons emerging from the collapse of asymmetric bubbles was studied 
in~\cite{Adib:2002ff} using effective radius as a measure. 
We study here the time evolution of the ellipticity via the quadrupole 
moment $\varrho_{ij}$ of the energy density $\rho(t,\boldsymbol{x})$
\begin{eqnarray}
\varrho_{ij} = \frac{ \int d^2 x \, x_i x_j \rho(t,\boldsymbol{x})}
{\int d^2 x \, \rho(t,\boldsymbol{x})}\,.
\label{quadrupole-moment}
\end{eqnarray}
Ellipticity is given by the eigenvalues of the traceless matrix
\begin{eqnarray}
\Theta_{ij} = \varrho_{ij} - \frac{1}{2}\delta_{ij}\varrho_{ij}\,.
\label{ellipticity}
\end{eqnarray}
Because the major and minor axis are along $x_1$ and $x_2$ the 
off-diagonal entries $\Theta_{12},\Theta_{21}$ vanish 
(a good check for the numerical accuracy of the method: 
$|\Theta_{12}|,|\Theta_{21}| <10^{-15}$) and 
then ellipticity is given directly by the diagonal elements 
of~(\ref{ellipticity}). 
We measured $\varrho_{ij}$ in a square of length 20 in physical units 
located in the center of 
the lattice. Lattice size was set to be $4800^2$ and thus no boundary 
effects will be important until far after time = 600. 
Figure~\ref{f:phi4ellipticity} 
shows $|\Theta_{11}|$ 
as a function of time when initially the ratio of the major to the 
minor axes was set to be $3:2$ (with the width in the $x_1$ direction 
being $r_0 = 3.9$). The oscillon approaches rapidly a 
spherical profile
and ellipticity disappears in exponential phases. 
The orientation of the major axis chances in a way that is not 
obviously periodic (zero crossings of $\Theta_{11}$ appearing in 
the Figure~\ref{f:phi4ellipticity} as almost vertical lines, which 
is demonstrated much more illustratively in Figure~\ref{f:phi4elliptical}). 
Figure~\ref{f:ellipticalenergyshells} shows total energy and energy 
inside two shells around the oscillon core. 
Though spherically symmetric shells are not an entirely adequate method to 
measure energy of an elliptic configuration, they give a reasonable estimate 
and show that while ellipticity disappears exponentially 
decreasing four orders of magnitude, the energy inside the oscillon core 
has decreased by just over 10\% indicating a quick collapse to a 
spherical shape. 

Figure~\ref{f:SGellipticity} shows the ellipticity 
and~\ref{f:ellipticalenergyshellsSG} total energy and energy inside shells
in a simulation with the same set-up but for sine-Gordon potential. 
Qualitatively features are similar, 
ellipticity decays rapidly and oscillon collapses into spherical profile. 
Here the orientation of major axis changes frequently in almost periodic way. 
Moreover, the decay of ellipticity $|\Theta_{11}|$ 
(in Figure~\ref{f:SGellipticity}) matches better to a power law 
than an exponential fit. It is very tempting to try to interpret this 
result on the basis of the obtained differences of oscillons in 
frequency space between quartic and sine-Gordon potentials. As both even and 
odd multiples of the oscillation frequncy are present in case of the quartic 
potential, the first radiative frequency, twice the basic frequency, 
has far greater amplitude than in sine-Gordon potential where the first 
radiative mode has stronger suppression its frequency being three 
times the basic oscillation frequency. Therefore it seems plausible that 
oscillon in quartic potential can radiate its asymmetry exponentially, 
while initial deviation from spherical symmetry decays only 
with a power law in sine-Gordon potential. 
Naturally, the study above does not exclude the 
possibility of different spherically symmetric forms for an oscillon.

\section{Colliding oscillons}

\begin{figure}
\begin{center}

\includegraphics[height=0.2\textheight]{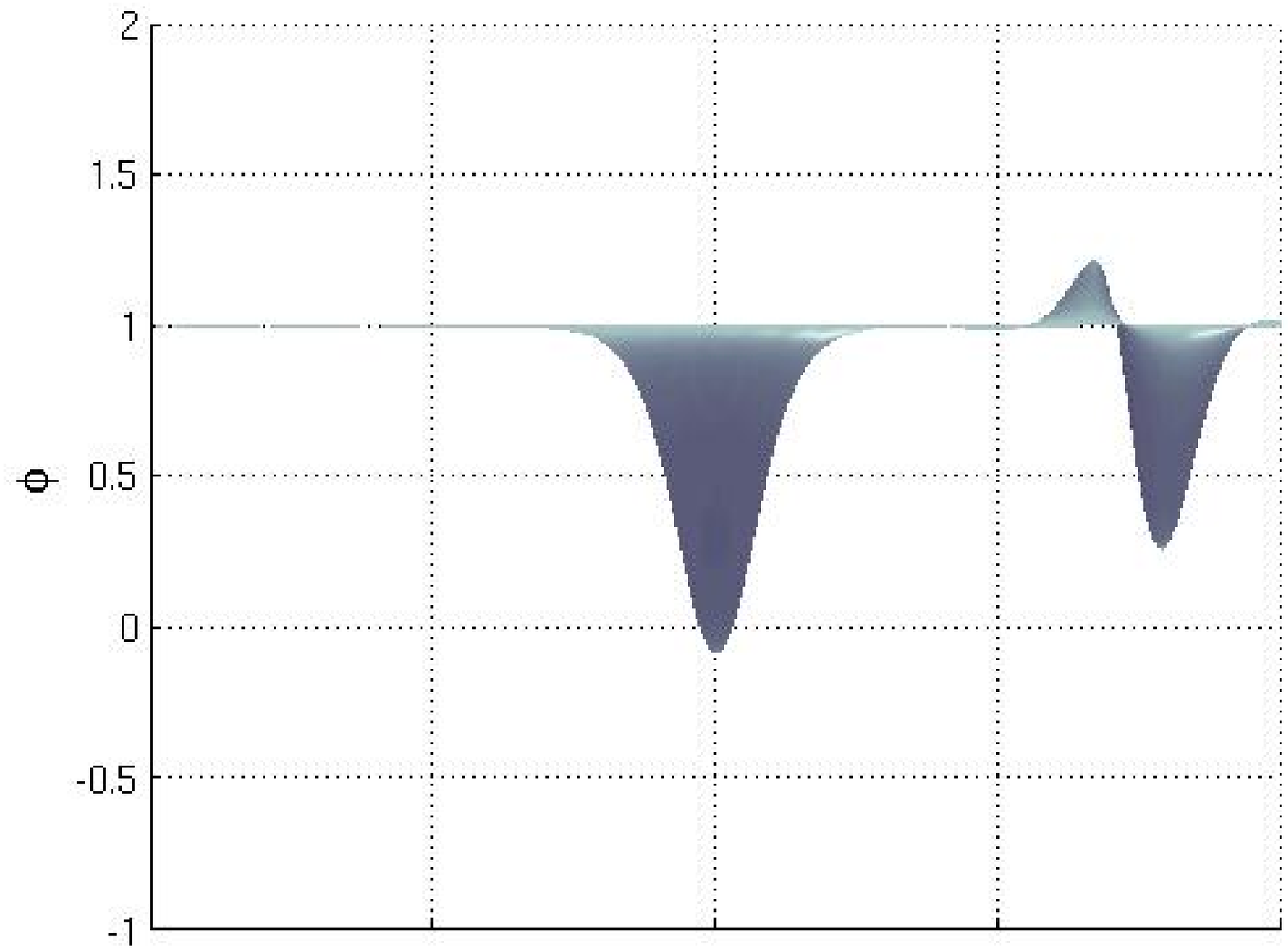}
\includegraphics[height=0.2\textheight]{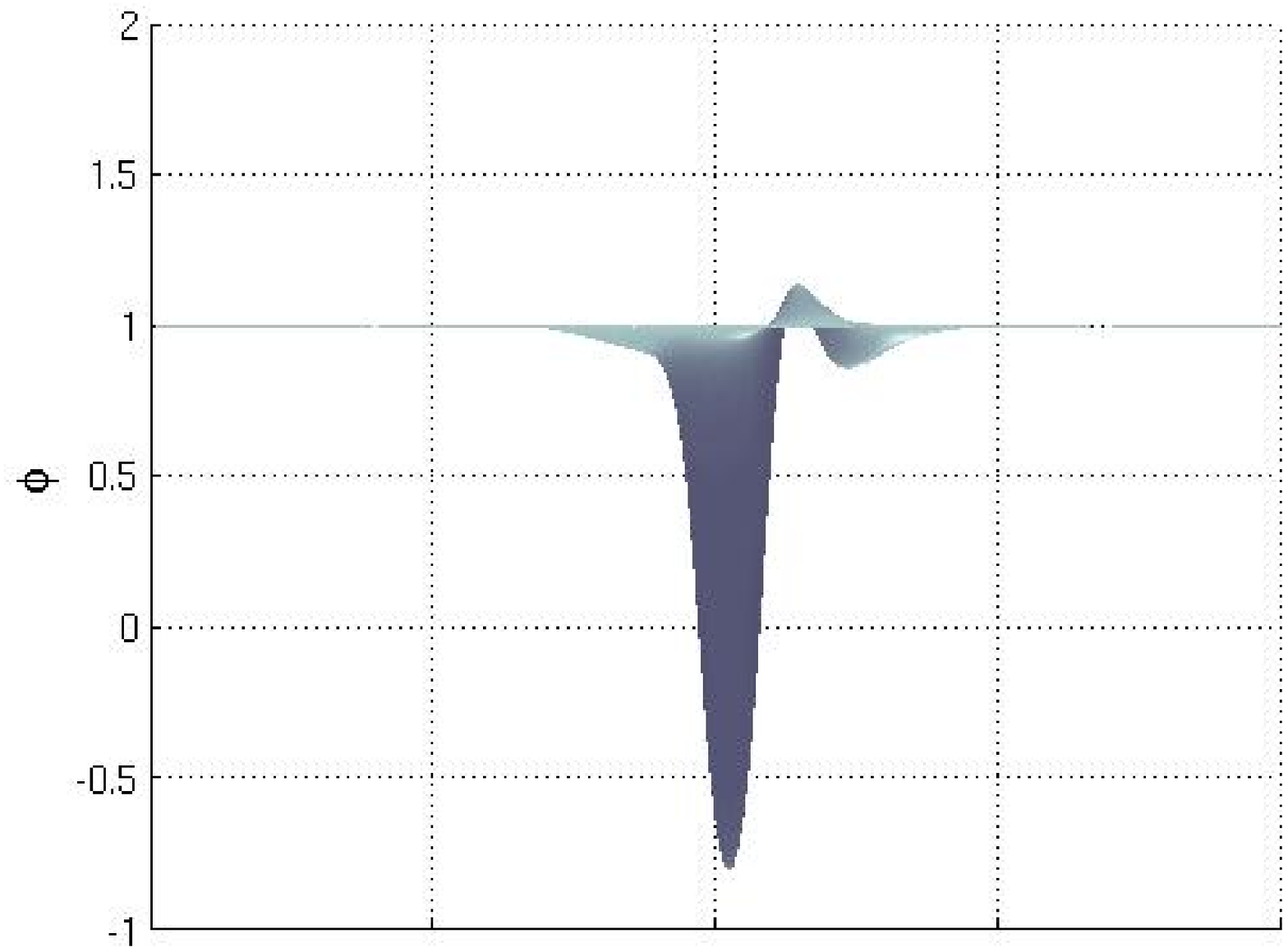}
\includegraphics[height=0.2\textheight]{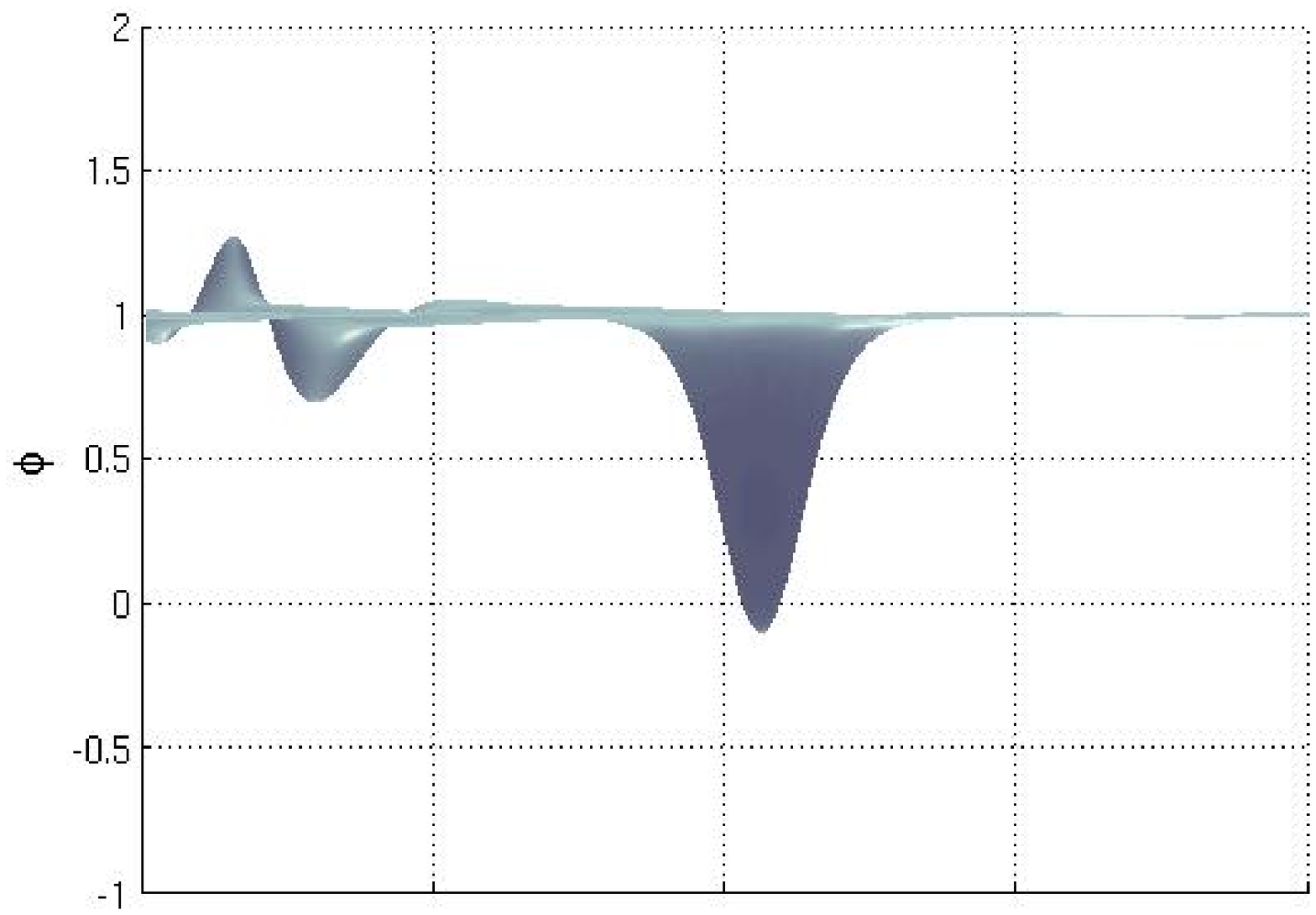}
  \caption{\label{f:phi4collisionOn} Sequence of snapshots at 
times $t= 0,\, 36.25,\, 80 $ of an on-axis collision of 
two oscillons, an immobile 
(originally located in the middle) and moving (from right to left), 
in the 2D $\phi^4$ theory.}

\end{center}
\end{figure}

\begin{figure}
\begin{center}

\includegraphics[height=0.2\textheight]{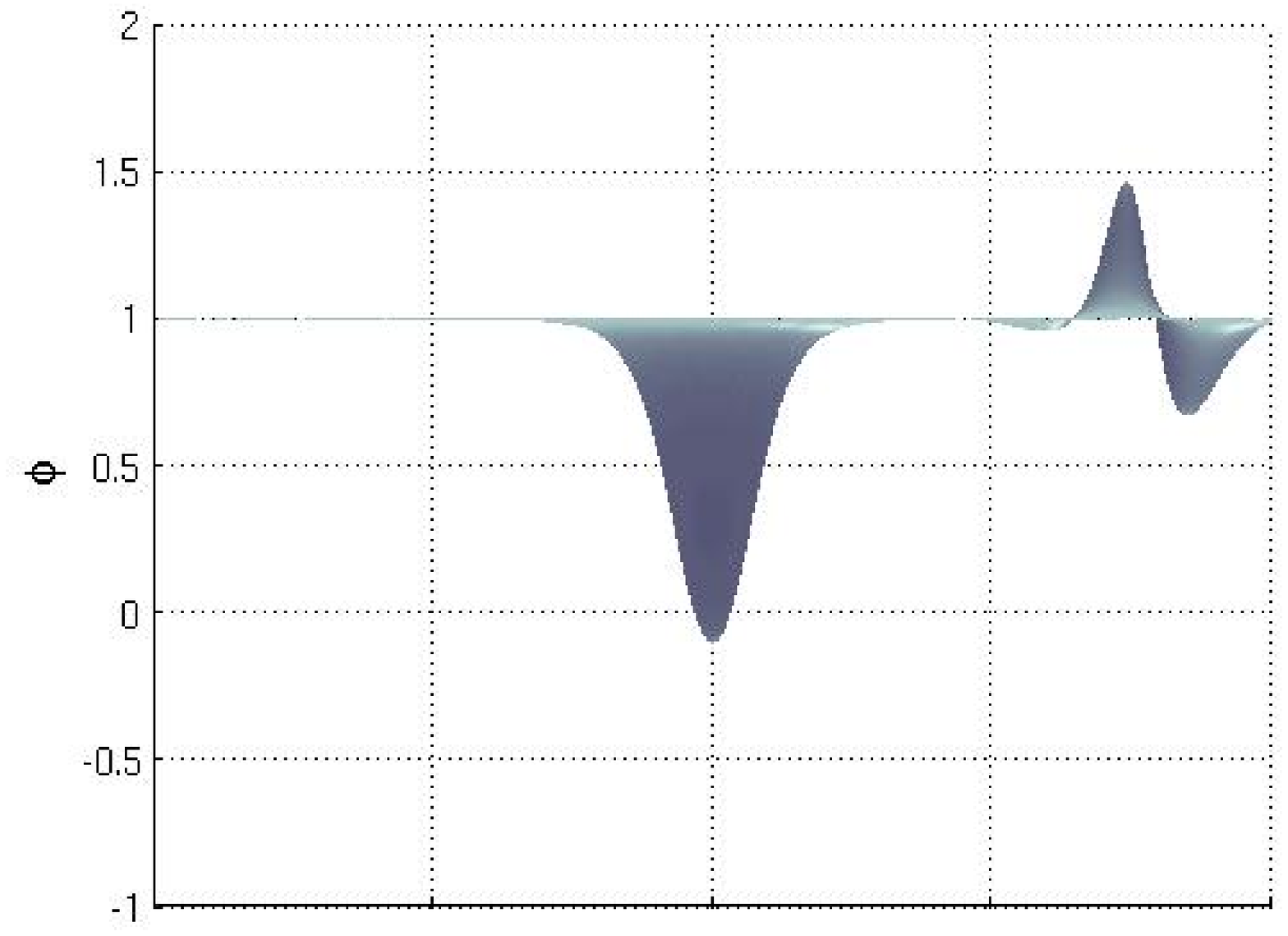}
\includegraphics[height=0.2\textheight]{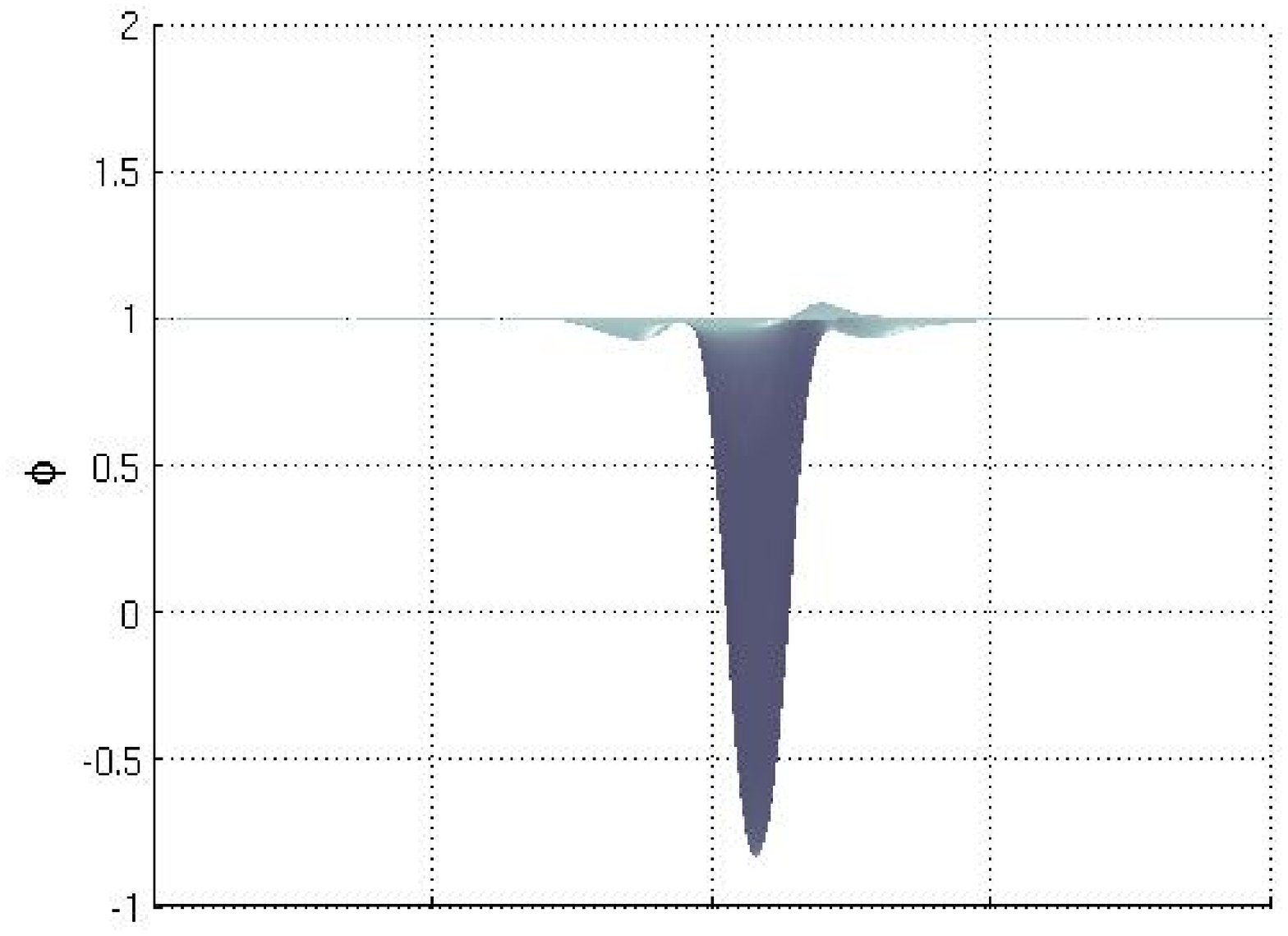}
\includegraphics[height=0.2\textheight]{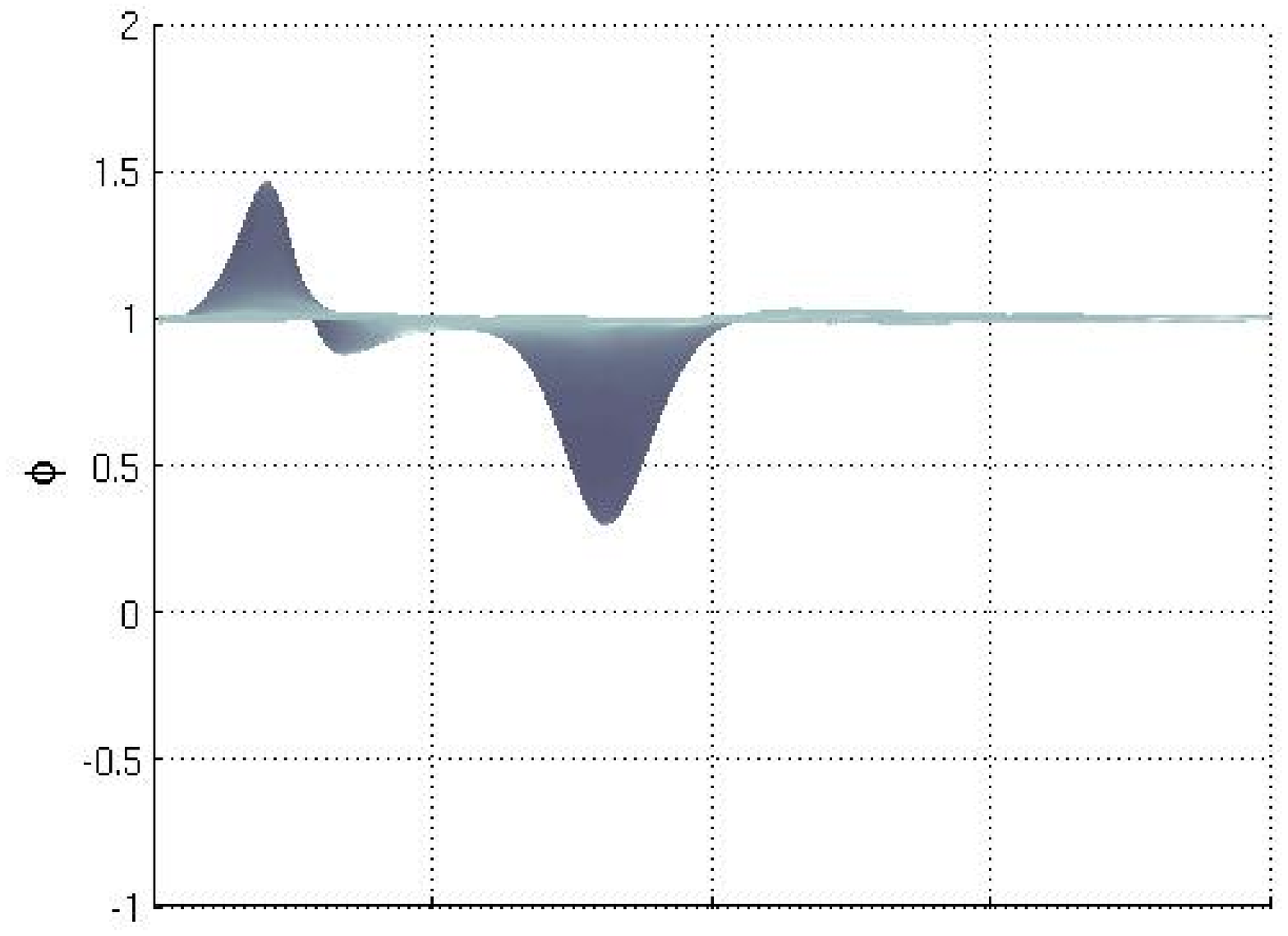}
  \caption{\label{f:phi4collisionOnb} 
Sequence of snapshots at times $t= 0,\, 36.75,\, 100$ of an on-axis 
  collision of two oscillons in the 2D $\phi^4$ theory. 
The phase at collision set  $1/4$ of the period different 
compared to Figure~\ref{f:phi4collisionOn}.
}

\end{center}
\end{figure}

\begin{figure}
\begin{center}

\includegraphics[height=0.2\textheight]{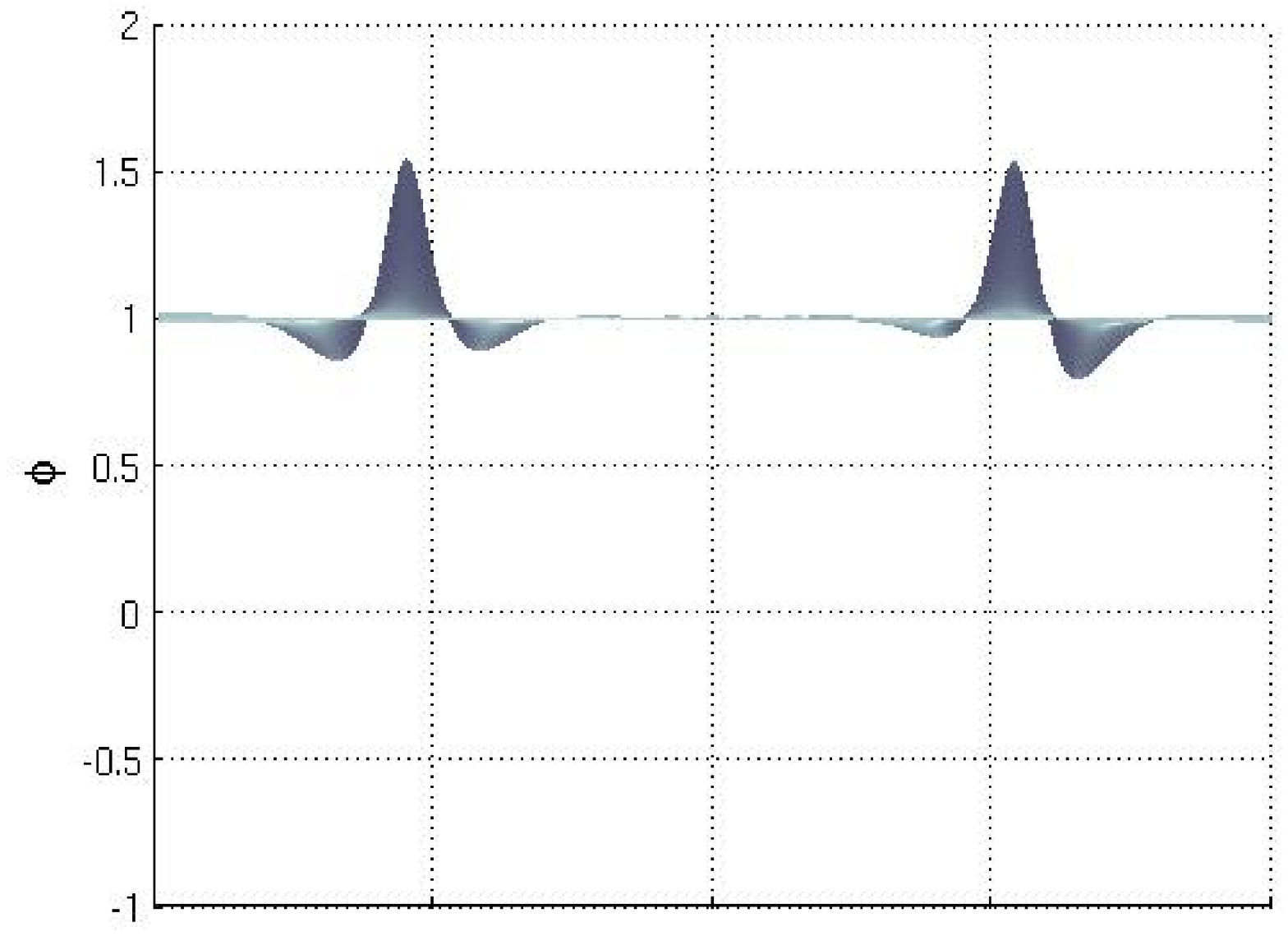}
\includegraphics[height=0.2\textheight]{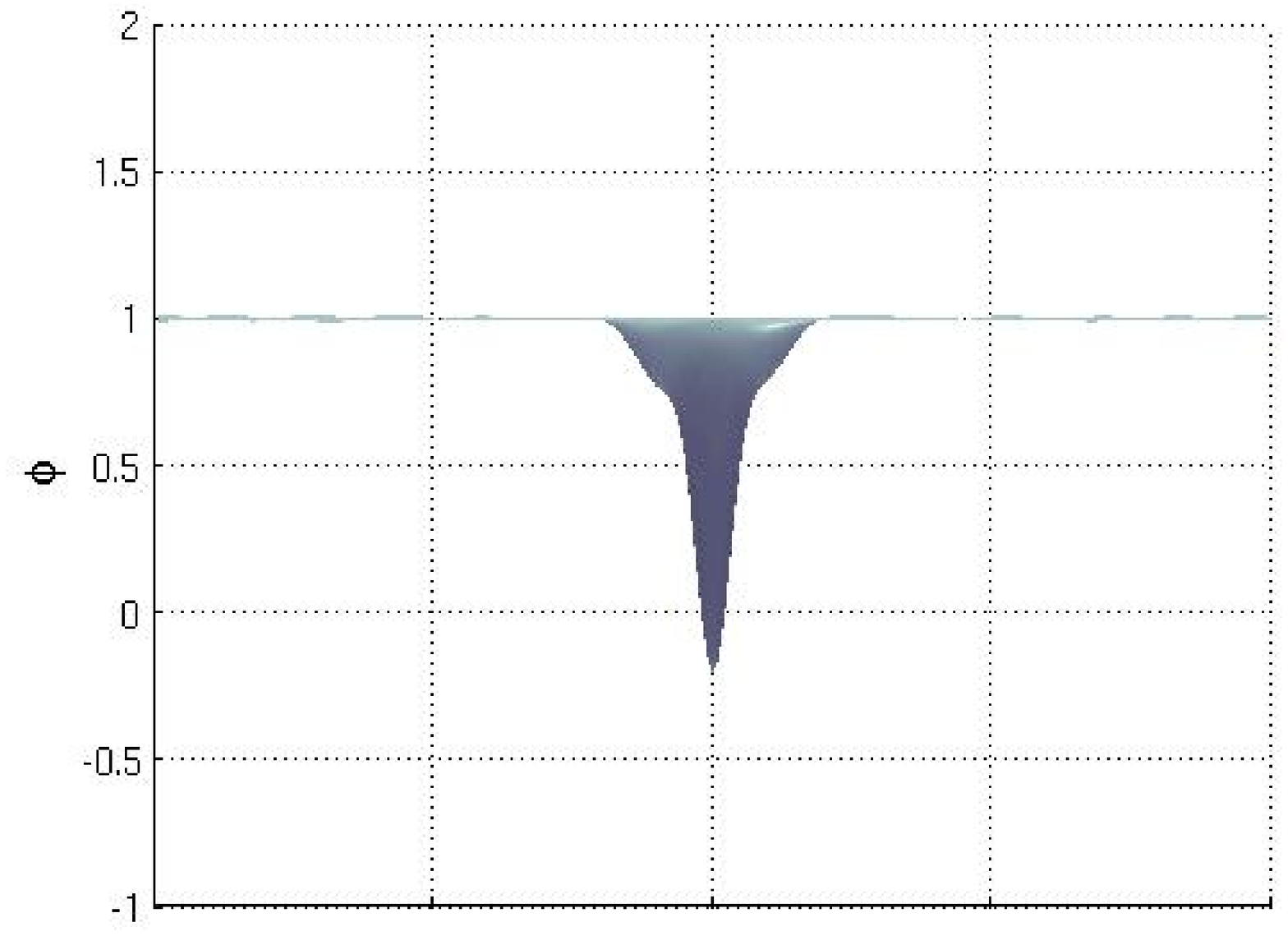}
\includegraphics[height=0.2\textheight]{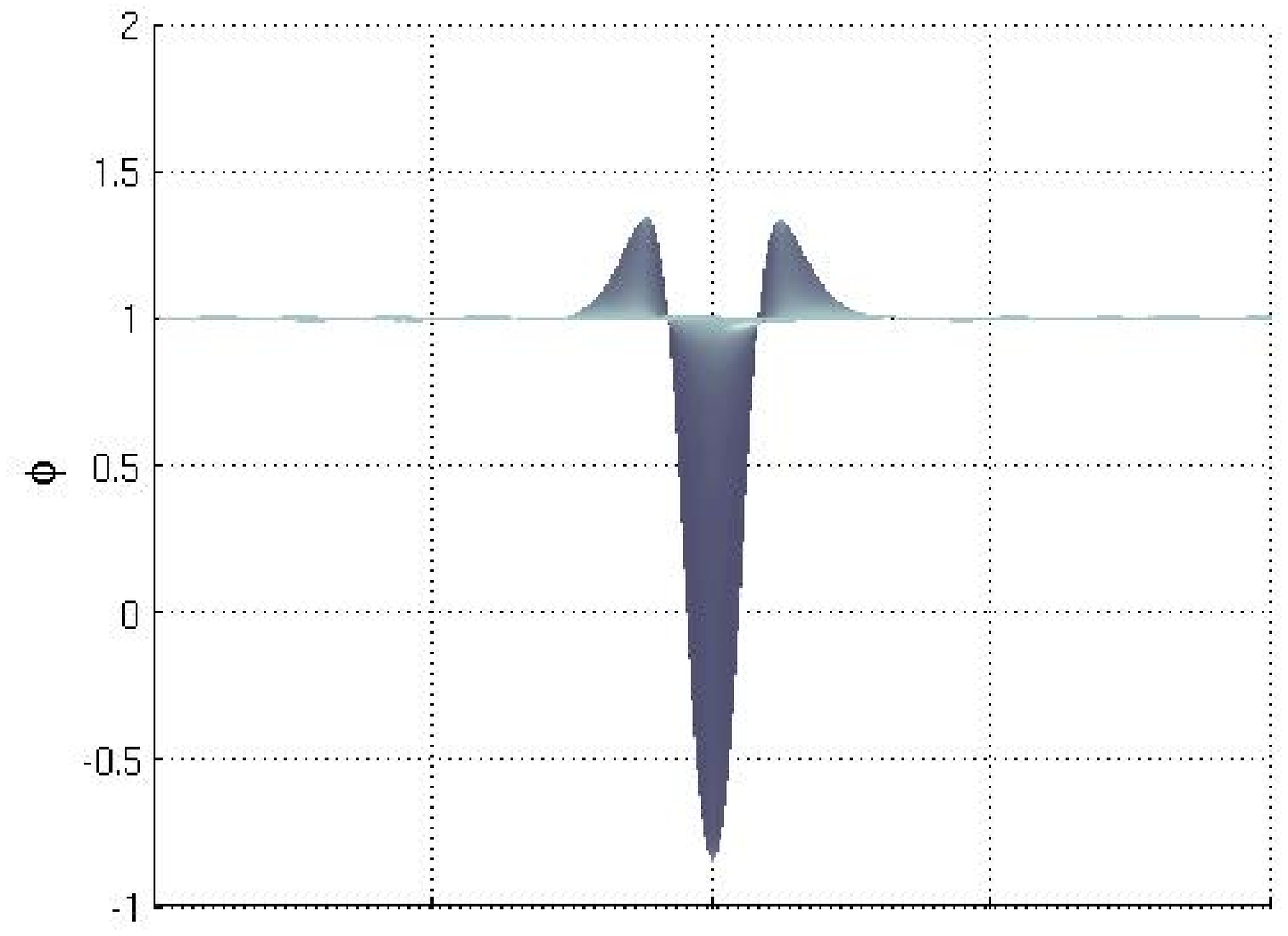}
\includegraphics[height=0.2\textheight]{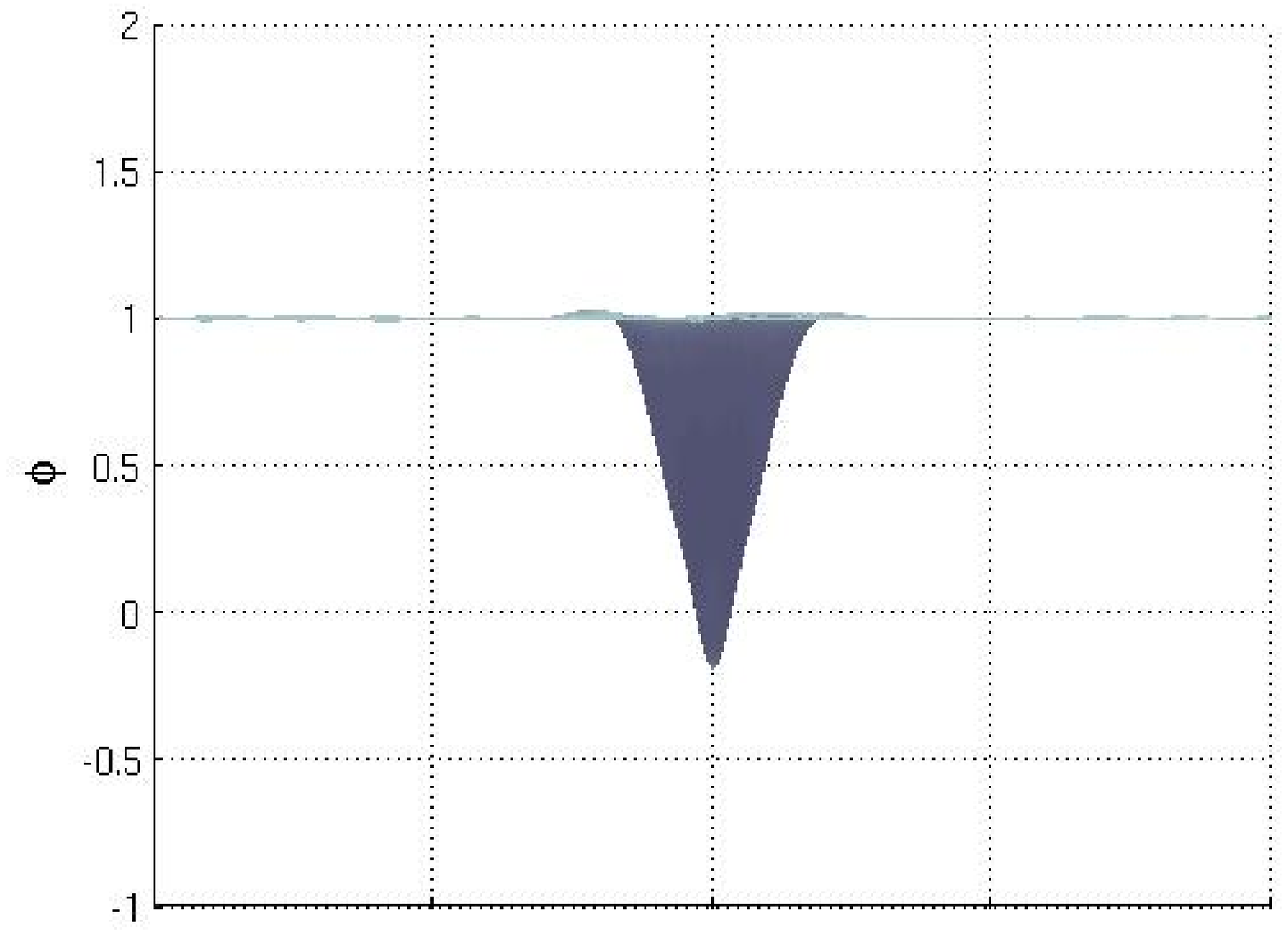}
\includegraphics[height=0.2\textheight]{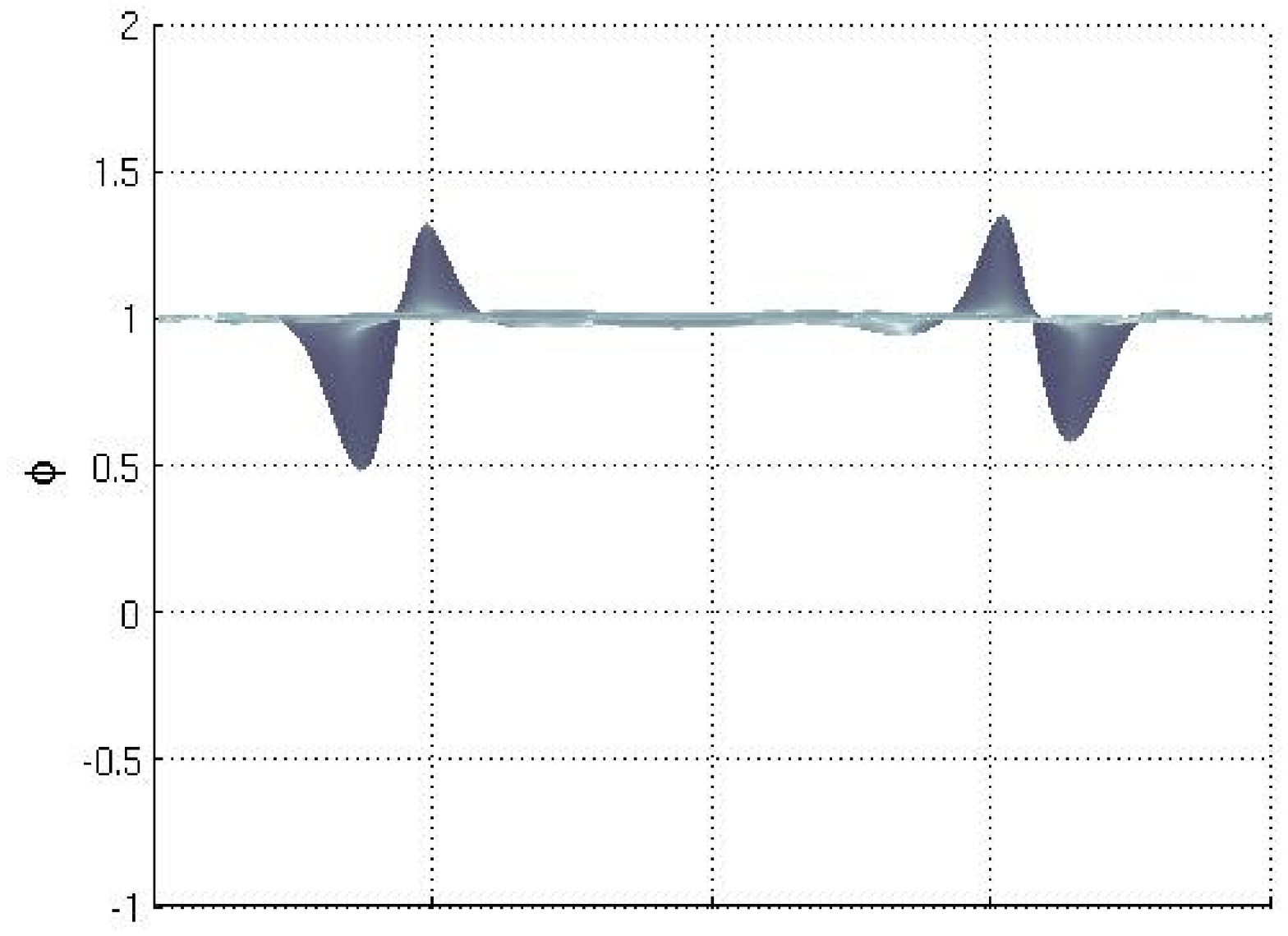}
\caption{\label{f:phi4collisionOn2} Sequence of snapshots at times 
$t= 0, \, 29.5,$ $ 30.5,\, 31.5,\, 60.5$ of an on-axis 
  collision of two oscillons in the 2D $\phi^4$ theory.}

\end{center}
\end{figure}

\begin{figure}
\begin{center}
\includegraphics[width=0.94\hsize]{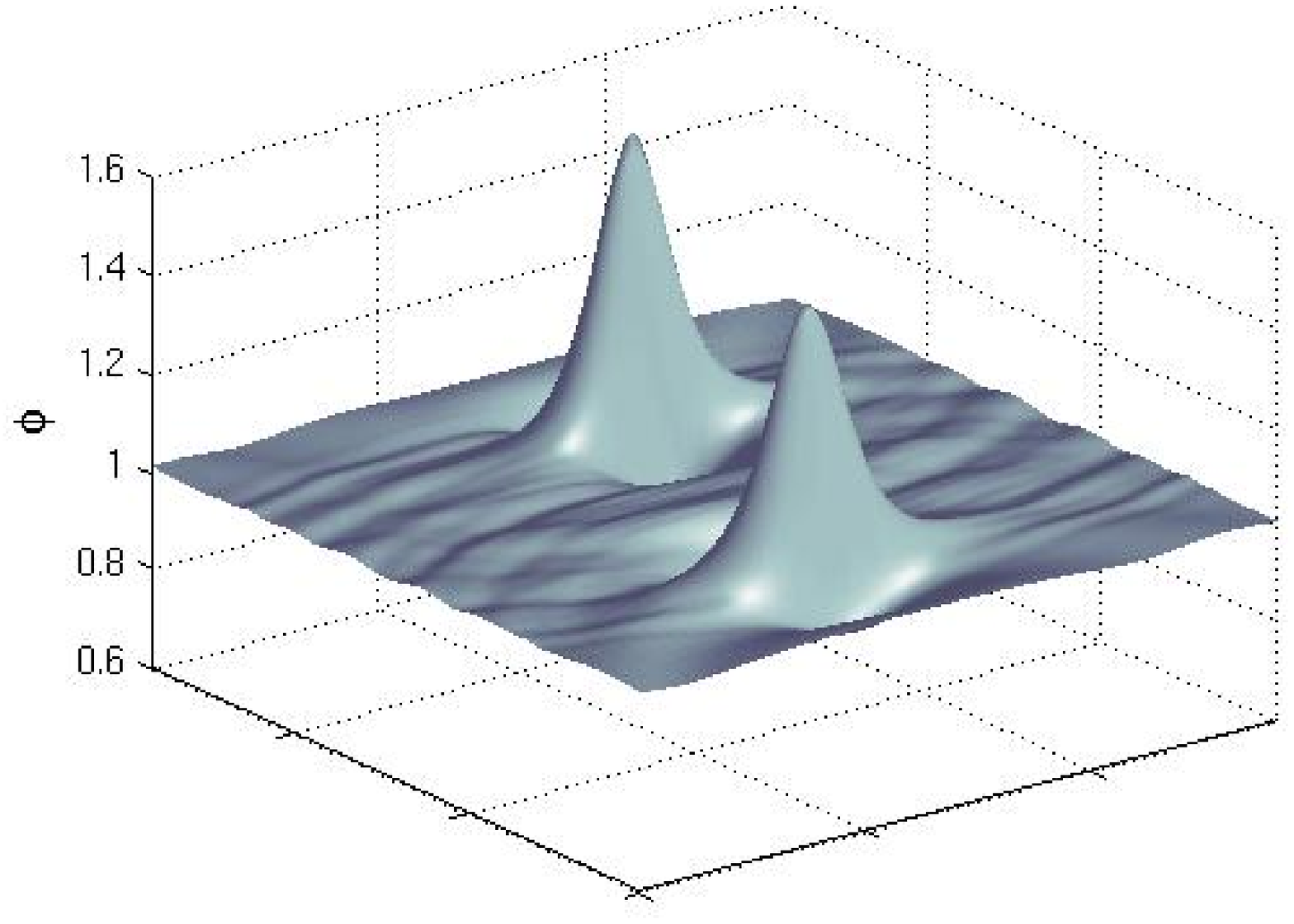}
\includegraphics[width=0.94\hsize]{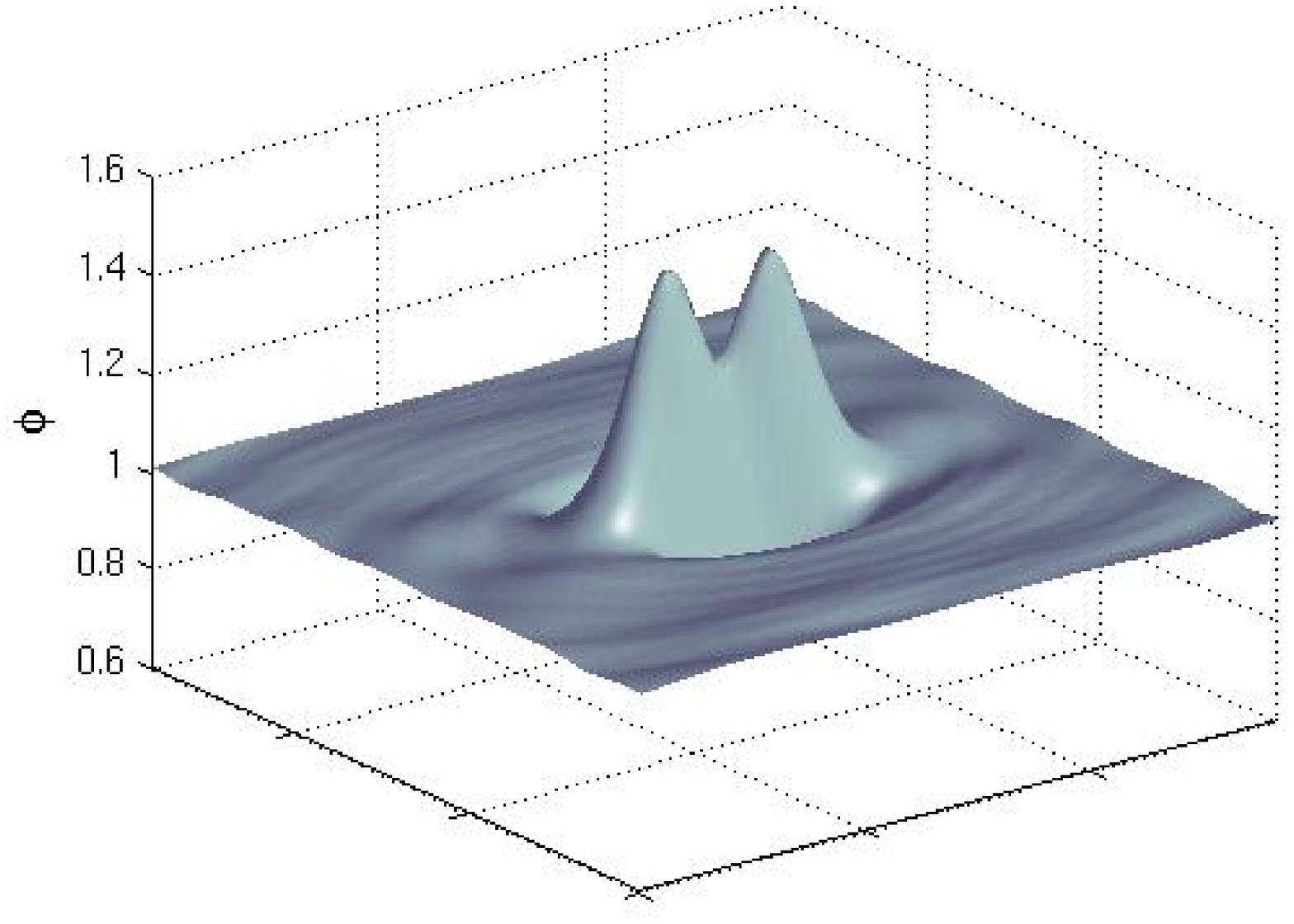}
\includegraphics[width=0.94\hsize]{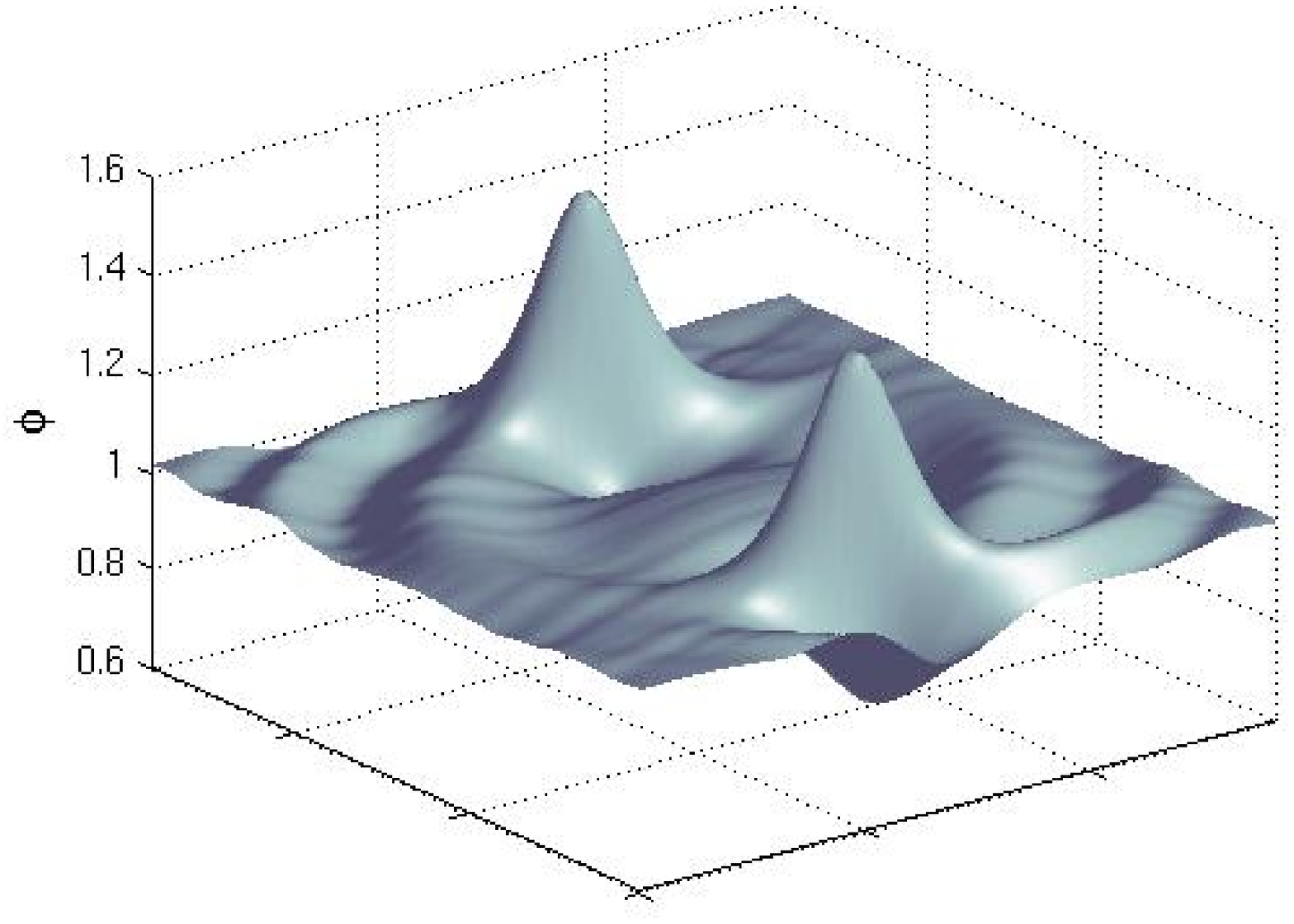}
  \caption{\label{f:phi4collisionOff} Sequence of snapshots 
at times $t=0,\,22,\,49$ of an 
 off-axis collision of two oscillons in the 2D $\phi^4$ theory.}
\end{center}
\end{figure}

If created in Nature via some mechanism like collapse of domains, 
oscillons would have initially translational momentum. 
In order to draw the connection between oscillons given birth 
in simulations with random initial conditions 
and stationary oscillons created with e.g. 
Gaussian ansatz, moving oscillons were prepared. 
Practically the oscillon that starts with a Gaussian initial 
profile is first allowed to 
evolve few oscillations and then the configuration is Lorentz boosted. 
For all the shown data the velocity of moving oscillons 
is set to be $0.5$ (we point out that oscillons originating 
from collapsing domains seem to generally have higher velocities).

Fast moving oscillons become some sort of waves: they have typically at least 
two 'phases', part of the wave front lies above, 
part below the minimum of the potential. 
There is a precise correspondance between boosted oscillons 
and the objects travelling on wave fronts originating from collapsing domains. 
We have also checked that boosted oscillon is persistent like a stationary one 
and can circulate on a lattice (with periodic boundary conditions) 
travelling long distances without demise. 
Here it is appealing to speculate that there is a connection to freak waves 
in the context of pattern formation in 
oceonography (for a study of freak wave formation from a nonlinear 
Schr\"odinger equation see~\cite{Onorato}).

We finish by reporting the results of studies where oscillons 
were made to collide. 
There is twofold reasoning for this study. 
We want to examine 
what kind of interactions 
oscillons have and thus try to understand their nature from 
that point of view. 
Secondly, if created in large numbers in phase transitions, 
collisions are inevitable and the behaviour of oscillons in those 
has thus impact on the subsequent evolution of the system.
The general rule is that a collision does not destroy an oscillon: we 
have never witnessed that the first collision between oscillons would have 
lead to a demise of either of them though it is noticeable that 
oscillons disturbed this way have also 
disappered in the second collision that 
occured on a lattice with periodic boundary conditions.

Figure~\ref{f:phi4collisionOn} shows snapshots from a collision between 
immobile and moving (velocity 0.5) oscillons. 
After the collision there is one fast moving oscillon continuing 
to the direction of original translational momentum and fairly 
stationary oscillon whose center, however, has moved slightly from the 
original location. This example seems thus almost interactionless. 
However, amount of interaction and momentum transfer depends on the 
phases of oscillons at the moment of collision and can result to 
two moving oscillons. 
This can be examined as the set-up for collision 
has been prepared by creating the 
immobile oscillon at arbitrary time on the lattice, we can control the 
phase of stationary oscillon via the moment of its creation.
The effect of the phase is demonstrated in 
Figure~\ref{f:phi4collisionOnb} 
where the set-up is otherwise the same as in 
Figure~\ref{f:phi4collisionOn}, but 
the initially stationary oscillon is time $1.3$ behind the previous 
one corresponding 
approximately $1/4$ of the period. Here both the oscillons move left 
after the collision. 

Figure~\ref{f:phi4collisionOn2} shows snapshots of a collison 
between two boosted oscillons with the same velocity (0.5) travelling into 
opposite directions. As can be seen oscillons pass through each other. 
We have also changed the alignment of oscillons so that the 
collision does not occur head on. Figure~\ref{f:phi4collisionOff} 
shows snapshots were the deviation in the alignment between the 
centers of oscillons is $5.0$ in physical units. 
The set-up leads to 
an attractive scattering, oscillons do not behave as classical 
particle like objects, but their paths bend towards each other the angle 
between initial and final velocity being approximately $20^{\circ}$.
%
\section{Conclusions}

We have studied oscillons, extremely long living oscillations of a 
scalar field, numerically in two dimensions for $\phi^4$ and 
sine-Gordon potentials.
We evolved stationary oscillon for $10^6$ time units or longer 
(which is three orders magnitude larger than the reported life time 
for I-balls, $10^3m^{-1}$, in~\cite{Kasuya:2002zs}) and examined 
the power spectrum of oscillations. 
Study in the frequency space not only suggested much longer life time 
on the basis of the time evolution of the basic oscillation frequency 
than we could directly probe, but also points out the validity of 
the separable ansatz~(\ref{solution-ansatz}). 
Though we have not evolved the oscillon in 
sine-Gordon potential as long, the analysis of the power spectrum 
suggests that it might be even more persistent than oscillons in $\phi^4$ 
theory in spite of the modulated amplitude. 
Studies of Lorentz boosted oscillons and collisions between them 
show also persistence of these objects and point 
out the behaviour that is very similar to solitons.

As our study was carried out on lattices with periodic boundary conditions, 
the radiation oscillons emit comes to some extent back to 
the centre of the lattice where oscillon is located. 
This can be viewed as a small perturbation and long survival of 
oscillons in the set-up as their persistence to radiation. 
On the other hand incoming 
radiation can also work otherwise pumping energy back into oscillon 
and thus extending the life time~\cite{Watkins}. 
Even in the latter case our study has relevance in situation where 
oscillon can absorb energy from enviroment, like a weak heat bath or a 
laboratory experiment in a closed system.

There are several unanswered question concerning oscillons 
and their longevity. The mechanism that makes some potentials 
able to confine energy over millions of oscillations should be 
understood together with the relation to the oscillation frequency.
On the basis of the spreading of the energy density during 
the oscillation (see Figure~\ref{f:energy-density}) one would 
naively assume that energy could escape in the form of 
radiation relatively quickly. The oscillation frequency also 
seem to be fined tuned just below the radiation frequency. 
This has been reported also for the SU(2) model oscillon~\cite{Farhi:2005rz} 
where the difference between oscillation and radiative 
frequencies was observed to be per mil level. 
Finally, even the level and role of nonlinearity in the 
nature of oscillons is not well understood. It has been noticed, 
both in the study of freak wave formation from a nonlinear 
Schr\"odinger equation~\cite{Onorato} as well as oscillon 
(called axiton by the authors) formation in the axion 
field~\cite{Kolb:1993hw}, 
that nonlinear effects are crucial in the generation of high 
density contrasts. 
On the other hand, in~\cite{Kasuya:2002zs} the authors emphasise 
the importance of the potential being dominated 
by the quadratic term as a condition for I-balls to exist 
as well as in~\cite{Watkins} weakness of nonlinear effects was 
considered important for the longevity of oscillons.

Almost stable, non-dissipative solutions of field theories 
oscillons have interest itself, but also their existence can lead 
to significant consequences that are worth studies of their own. 
In particular, if formed, oscillons could play 
a role in the early universe and phase transitions that occured there 
especially if they survive long enough in realistic conditions, 
as for instance their behaviour in collisions suggests. 
Here e.g. the oscillon in sine-Gordon potential may have relevance 
in the QCD phase transition as the axion potential has sine-Gordon form. 
Kolb and Tkachev have studied the formation of solitons 
in~\cite{Kolb:1993hw}. 
While finishing this paper we became aware of the 
study~\cite{Graham:2006xs} where similar life times of oscillons as we 
obtained here were reported, but in an expanding background 
in one dimension with a potential including also $\phi ^6$~-term.
If this longevity persists in full three dimensional models 
with Hubble expansion, 
oscillons may have interesting cosmological implications.

{\it Note added:} while this paper was being considered for publication 
the study~ \cite{Fodor:2006zs} appeared examining oscillons 
in 3 dimensions.


\begin{acknowledgments}

The authors thank Nick Manton, Bernard Piette and 
Anders Tranberg for discussions and Paul Saffin for pointing 
out the reference~\cite{Watkins}.
P. S. was supported by Marie Curie Fellowship of the
European Community Program HUMAN POTENTIAL under contract
HPMT-CT-2000-00096 and by the Netherlands
Organization for Scientific Research (N.W.O.) under the VICI
programme and 
acknowledges the hospitality at the University of Sussex.
 
\end{acknowledgments}


\bibliography{references}

\end{document}